\documentclass{sigplanconf}

\usepackage{times}
\usepackage{amsmath,amssymb, latexsym}

\usepackage[latin1]{inputenc}
\usepackage{amsmath}
\usepackage{amsfonts}
\usepackage{amssymb}
\usepackage{amsthm}

\usepackage[inference]{semantic}
\usepackage{enumerate}
\def\url{}
\usepackage{xspace}
\usepackage{epsfig}
\usepackage{mathpartir}
\usepackage{booktabs}
\usepackage{paralist}
\usepackage{hyperref}

\def\@envspa{\hspace{0.3em}}
\def\@sa{\hspace{-0.2em}}
\def\@sb{\hspace{0.5em}}
\def\@sc{\hspace{-0.1em}}

{\itshape}{}

\newtheorem*{lemma*}{Lemma}
\newtheorem{definition}{Definition}

\newtheorem*{theorem*}{Theorem}

\usepackage{color}
\usepackage{textcomp}
\usepackage{soul}

\definecolor{colorNV}{rgb}{1,0.8,1}

\definecolor{colorRJ}{rgb}{0.2,1.0,0.3}

\definecolor{colorAB}{rgb}{0.8,0.8,0}

\usepackage{commands}
\usepackage{liquidHaskell}
\usepackage{listings}

\def\withcolor{}

\ifdefined\withcolor
	 \definecolor{haskellblue}{rgb}{0.0, 0.0, 1.0}
	 \definecolor{haskellstr}{rgb}{0.2, 0.2, 0.6}
	 \definecolor{haskellred}{rgb}{1.0, 0.0, 0.0}
  \definecolor{gray_ulisses}{gray}{0.55}
  \definecolor{castanho_ulisses}{rgb}{0.71,0.33,0.14}
  \definecolor{preto_ulisses}{rgb}{0.41,0.20,0.04}
  \definecolor{green_ulises}{rgb}{0.2,0.75,0}
\else
	\definecolor{haskellblue}{gray}{0.1}
	\definecolor{haskellstr}{gray}{0.1}
	\definecolor{haskellred}{gray}{0.1}
	\definecolor{gray_ulisses}{gray}{0.1}
	\definecolor{castanho_ulisses}{gray}{0.1}
	\definecolor{preto_ulisses}{gray}{0.1}
	\definecolor{green_ulisses}{gray}{0.1}
\fi

\def\codesize{\small}

\lstdefinelanguage{HaskellUlisses} {
	basicstyle=\ttfamily\codesize,
	sensitive=true,
	morecomment=[l][\color{gray_ulisses}\ttfamily\codesize]{--},
	morecomment=[s][\color{gray_ulisses}\ttfamily\codesize]{\{-}{-\}},
	morestring=[b]",
  escapeinside={(*}{*)},
	stringstyle=\color{haskellstr},
	showstringspaces=false,
	numberstyle=\codesize,
	numberblanklines=true,
	showspaces=false,
	breaklines=true,
	showtabs=false,
  literate={<=}{{$\leq$}}1
           {>>=}{{>>=}}3
           {>=}{{$\geq$}}1
           {\\}{{$\lambda$}}1
           {!=}{{$\neq$}}1
           {forall}{{$\forall$}}1
           {->}{{$\rightarrow$}}2
           {Set_mem}{{$\in$}}1
           {Set_cup}{{$\cup$}}1
           {Set_cap}{{$\cap$}}1
           {Set_emp}{{$\emptyset$}}1
           {Set_sub}{{$\subseteq$}}1
           {<=>}{{$\Leftrightarrow$}}3
           {=>}{{$\Rightarrow$}}2
           {||-}{{$\vdash$}}1
           {|->}{{$\mapsto$}}2
           {<:}{{$\preceq$}}1
           {Inarritu}{Inarritu}8,  
	emph=
	{[1]
		FilePath,IOError,abs,acos,acosh,all,and,any,appendFile,approxRational,asTypeOf,asin,
		asinh,atan,atan2,atanh,basicIORun,break,catch,ceiling,chr,compare,concat,concatMap,
		const,cos,cosh,curry,cycle,decodeFloat,denominator,digitToInt,div,divMod,drop,
		dropWhile,either,elem,encodeFloat,enumFrom,enumFromThen,enumFromThenTo,enumFromTo,
		error,even,exp,exponent,fail,mapMaybe,filter,flip,floatDigits,floatRadix,floatRange,floor,
		fmap,foldl,foldl1,foldr,foldr1,fromDouble,fromEnum,fromInt,fromInteger,fromIntegral,
		fromRational,fst,gcd,getChar,getContents,getLine,head,id,inRange,index,init,intToDigit,
		interact,ioError,isAlpha,isAlphaNum,isAscii,isControl,isDenormalized,isDigit,isHexDigit,
		isIEEE,isInfinite,isLower,isNaN,isNegativeZero,isOctDigit,isPrint,isSpace,isUpper,iterate,
		last,lcm,length,lex,lexDigits,lexLitChar,lines,log,logBase,lookup,map,mapM,mapM_,max,
		maxBound,posMax,negMax,maximum,maybe,min,minBound,minimum,mod,negate,not,notElem,null,numerator,odd,
		or,ord,pi,pred,primExitWith,print,product,properFraction,putChar,putStr,putStrLn,quot,
		quotRem,range,rangeSize,read,readDec,readFile,readFloat,readHex,readIO,readInt,readList,readLitChar,
		readLn,readOct,readParen,readSigned,reads,readsPrec,realToFrac,recip,rem,repeat,replicate,return,
		reverse,round,scaleFloat,scanl,scanl1,scanr,scanr1,seq,sequence,sequence_,show,showChar,showInt,
		showList,showLitChar,showParen,showSigned,showString,shows,showsPrec,significand,signum,sin,
		sinh,snd,span,splitAt,sqrt,subtract,succ,sum,tail,take,takeWhile,tan,tanh,threadToIOResult,toEnum,
		toInt,toInteger,toLower,toRational,toUpper,truncate,uncurry,undefined,unlines,until,unwords,unzip,
		unzip3,userError,words,writeFile,zip,zip3,zipWith,zipWith3,listArray,doParse,empty,for,initTo,
        assert,compose,checkGE,maxEvens,empty,create,get,set,initialize,idVec,fastFib,fibMemo,
        ex1,ex2,ex3,incr,inc,dec,isPos,positives,find,insert,len,size,union,fromList,initUpto,trim,
        insertSort,decsort,qsort,reverse,append,upperCase, ifM, whileM, get, decrM, diff, 
        project, select, leq, elts, keys, dkeys, dfun, addKey, pTrue, emptyRD, rFalse,
        	dom, rng, isI, isD, isS, movie1, movie2,  toI, toS, toD, good_titles, runState, ret, 
        	update, getCtr, setCtr, ctr, rdCtr, wrCtr, ifTest, whileTest, posCtr, zeroCtr, decr, decCtr, 
        	pread , pwrite , plookup , pcontents, pcreateF , pcreateFP, pcreateD, active, caps, pset, eqP, 
        	write, contents, alloc, derivP, copyP, createDir, store, copyRec, copySpec, 
        	forM_, when, flookup, fread, createDir, pcreateFile, isFile, copyFrame
	},
	emphstyle={[1]\color{haskellblue}},
	emph=
	{[2] 	Show,Eq,Ord,Num,UpClosed,Comp,Wit,Witness,Inductive,Meet,Flip,TRUE,Nat,Pos,Neg,IntGE,Plus,List,
        Bool,Char,Double,Either,Float,IO,Integer,Int,Maybe,
        Ordering,Rational,Ratio,ReadS,ShowS,String,Word8,
        InPacket,Tree,Vec,NullTerm,IncrList,DecrList,
        UniqList,BST,MinHeap,MaxHeap,World,RIO,IO,HIO,Post,Pre,
        Privilege, Prop, Chain, ChainTy, Range, Dict, RD, Dom, Set, P, Univ, Schema, MovieSchema, RT, 
        TDom, TRange, MoviesTable, RTSubEqFlds, RTEqFlds, Disjoint, Union, Ret, Seq, Trans, Map, 
        Pure, Then, Else, Exit, Inv, OneState, Priv, Path, FH, Stable
	},
	emphstyle={[2]\color{castanho_ulisses}},
	emph=
	{[3]
		case,class,data,deriving,do,else,if,import,in,infixl,infixr,instance,let,
		module,measure,of,primitive,then,refinement,type,where,forall,bound, otherwise
	},
	emphstyle={[3]\color{preto_ulisses}\textbf},
	emph=
	{[4]
		quot,rem,div,mod,elem,notElem,seq
	},
	emphstyle={[4]\color{castanho_ulisses}\textbf},
	emph=
	{[5]
		EQ,False,GT,Just,LT,Left,Nothing,Right,True, D, I, S
	},
	emphstyle={[5]\color{preto_ulisses}\textbf}
}

\lstnewenvironment{code}
{\lstset{language=HaskellUlisses}}
{}

\lstnewenvironment{mcode}
{\lstset{language=HaskellUlisses,mathescape}}
{}

\lstMakeShortInline[language=HaskellUlisses,basicstyle=\ttfamily\normalsize,breakatwhitespace]@

\lstdefinelanguage{Pseudo} {
	basicstyle=\ttfamily\codesize,
	sensitive=true,
  mathescape=true,
	morecomment=[l][\color{gray_ulisses}\ttfamily\codesize]{--},
	morecomment=[s][\color{gray_ulisses}\ttfamily\codesize]{\{-}{-\}},
	morestring=[b]",
	showstringspaces=false,
	numberstyle=\codesize,
	numberblanklines=true,
	showspaces=false,
	breaklines=true,
	showtabs=false
}

\usepackage{flushend}

\usepackage{inconsolata}

\usepackage{thmtools}
\declaretheoremstyle[%
  spaceabove=-6pt,%
  spacebelow=6pt,%
  headfont=\normalfont\itshape,%
  postheadspace=1em,%
  qed=\qedsymbol,%
  headpunct={}
]{mystyle}

\newcommand\ifextended[2]{#2}
\conferenceinfo{CONF 'yy}{Month d--d, 20yy, City, ST, Country} 
\copyrightyear{20yy} 
\copyrightdata{978-1-nnnn-nnnn-n/yy/mm} 
\doi{nnnnnnn.nnnnnnn}

\titlebanner{banner above paper title}        
\preprintfooter{short description of paper}   

\title{Bounded Refinement Types
\thanks{This work was supported by NSF grants CCF-1422471, C1223850, CCF-1218344,
        a Microsoft Research Ph.D Fellowship and a generous gift from Microsoft
        Research.}\ifextended{\\ Supplementary Material}{}}

\authorinfo{Niki Vazou \and Alexander Bakst \and Ranjit Jhala}{UC San Diego}{}

\renewcommand{\keywords}{\paragraph*{Keywords}}

\begin{document}
\toappear{}
\maketitle

\begin{abstract}
We present a notion of bounded quantification for refinement types
and show how it expands the expressiveness of refinement typing
by using it to develop typed combinators for:
(1)~relational algebra and safe database access,
(2)~Floyd-Hoare logic within a state transformer
    monad equipped with combinators for branching
    and looping, and
(3)~using the above to implement a refined IO
    monad that tracks capabilities and resource usage.
This leap in expressiveness comes via a translation to ``ghost" functions,
which lets us retain the automated and decidable SMT based checking and
inference that makes refinement typing effective in practice.

\end{abstract}


\keywords
haskell, refinement types, abstract interpretation

\section{Introduction} \label{sec:intro}

Must program verifiers always choose between expressiveness
and automation?
On the one hand, tools based on higher order logics
and full dependent types impose no limits on expressiveness,
but require user-provided (perhaps, tactic-based) proofs.
On the other hand, tools based on Refinement Types~\cite{Rushby98,pfenningxi98}
trade expressiveness for automation. For example, the refinement types
\begin{code}
  type Pos     = {v:Int | 0 < v}
  type IntGE x = {v:Int | x <= v}
\end{code}
specify subsets of @Int@ corresponding to values
that are positive or larger than some other value @x@
respectively. By limiting the refinement predicates to
SMT-decidable logics~\cite{NelsonOppen}, refinement type
based verifiers eliminate the need for explicit proof terms,
and thus automate verification.


This high degree of automation has enabled the
use of refinement types for a variety of verification
tasks, ranging from array bounds checking~\cite{LiquidPLDI08},
termination and totality checking~\cite{LiquidICFP14},
protocol validation~\cite{GordonTOPLAS2011,FournetCCS11},
and securing web applications~\cite{SwamyOAKLAND11}.
Unfortunately, this automation comes at a price.
To ensure predictable and decidable type checking, we must
limit the logical formulas appearing in specification types
to decidable (typically quantifier free) first order theories,
thereby precluding \emph{higher-order} specifications that
are essential for \emph{modular} verification.

In this paper, we introduce \emph{Bounded Refinement Types}
which reconcile expressive higher order specifications
with automatic SMT based verification. Our approach
comprises two key ingredients.
Our first ingredient is a mechanism, developed by \cite{vazou13},
for \emph{abstracting} refinements over type signatures.
This mechanism is the analogue of parametric polymorphism
in the refinement setting: it increases expressiveness by
permitting generic signatures that are universally quantified
over the (concrete) refinements that hold at different
call-sites.
However, we observe that for modular verification, we
additionally need to \emph{constrain} the abstract refinement
parameters, typically to specify fine grained dependencies
between the parameters.
Our second ingredient provides a technique
for enriching function signatures with \emph{subtyping constraints}
(or \emph{bounds}) between abstract refinements that must be
satisfied by the concrete refinements at instantiation.
Thus, constrained abstract refinements are the analogue of bounded
quantification in the refinement setting and in this paper, we
show that this simple technique proves to be remarkably effective.


\begin{itemize}
\item
First, we demonstrate via a series of short examples how bounded refinements
enable the specification and verification of diverse textbook higher order
abstractions that were hitherto beyond the scope of decidable refinement
typing~(\S~\ref{sec:overview}).

\item
Second, we formalize bounded types and show how bounds are translated
into ``ghost'' functions, reducing type checking and inference to the
``unbounded'' setting of~\cite{vazou13}, thereby ensuring that checking
remains decidable. Furthermore, as the bounds are Horn constraints, we
can directly reuse the abstract interpretation of Liquid Typing~\citep{LiquidPLDI08}
to automatically infer concrete refinements at instantiation
sites~(\S~\ref{sec:check}).

\item
Third, to demonstrate the expressiveness of bounded refinements, we
use them to build a typed library for extensible dictionaries, to
then implement a relational algebra library on top of those
dictionaries, and to finally build a library for type-safe
database access~(\S~\ref{sec:database}).

\item
Finally, we use bounded refinements to develop a \emph{Refined State Transformer}
monad for stateful functional programming, based upon Filli\^atre's method
for indexing the monad with pre- and post-conditions~\citep{Filliatre98}.
We use bounds to develop branching and looping combinators whose types
signatures capture the derivation rules of Floyd-Hoare logic, thereby
obtaining a library for writing verified stateful computations~(\S~\ref{sec:state}).
We use this library to develop a refined IO monad that tracks capabilities
at a fine-granularity, ensuring that functions only access specified
resources~(\S~\ref{sec:files}).
\end{itemize}

We have implemented Bounded Refinement Types in \toolname~\citep{LiquidICFP14}.
The source code of the examples (with slightly more verbose concrete syntax)
is at \cite{liquidhaskellgithub}.
While the construction of these verified abstractions is possible with full
dependent types, bounded refinements
keep checking automatic and decidable,
use abstract interpretation to automatically synthesize
refinements (\ie pre- and post-conditions and loop invariants),
and most importantly
enable retroactive or \emph{gradual} verification as when
erase the refinements, we get valid programs in the
host language~(\S~\ref{sec:related}).
%
%
Thus, bounded refinements point a way towards keeping our automation, and
perhaps having expressiveness too.
%

\section{Overview}\label{sec:overview}

We start with a high level overview of bounded refinement types.
To make the paper self contained, we begin by recalling the notions
of abstract refinement types. Next, we introduce bounded refinements,
and show how they permit \emph{modular} higher-order specifications.
Finally, we describe how they are implemented via an elaboration
process that permits \emph{automatic} first-order verification.

\subsection{Preliminaries}

\paragraph{Refinement Types} let us precisely specify subsets of values,
by conjoining base types with logical predicates that constrain the values.
We get decidability of type checking, by limiting these predicates to
decidable, quantifier-free, first-order logics, including the theory
of linear arithmetic, uninterpreted functions, arrays, bit-vectors
and so on. Apart from subsets of values, like the @Pos@ and @IntGE@
that we saw in the introduction, we can specify contracts
like pre- and post-conditions by suitably refining the input
and output types of functions.

\paragraph{Preconditions} are specified by refining input types.
We specify that the function @assert@ must
\emph{only} be called with @True@,
where the refinement type @TRUE@ contains only the singleton @True@:
\begin{code}
  type TRUE = {v:Bool | v <=> True}

  assert         :: TRUE -> a -> a
  assert True x  = x
  assert False _ = error "Provably Dead Code"
\end{code}

\paragraph{We can specify post-conditions} by refining output types.
For example, a primitive @Int@ comparison operator @leq@ can be
assigned a type that says that the output is @True@ iff the
first input is actually less than or equal to the second:
\begin{code}
  leq :: x:Int -> y:Int -> {v:Bool | v <=> x <= y}
\end{code}

\paragraph{Refinement Type Checking} proceeds by checking that at each
application, the types of the actual arguments are \emph{subtypes}
of those of the function inputs, in the environment (or context) in
which the call occurs.
Consider the function:
\begin{code}
  checkGE     :: a:Int -> b:IntGE a -> Int
  checkGE a b = assert cmp b
    where cmp = a `leq` b
\end{code}
To verify the call to @assert@ we check that
the actual parameter @cmp@ is a subtype of @TRUE@,
under the assumptions given by the input types for
@a@ and @b@.
Via subtyping~\cite{LiquidICFP14} the check reduces to establishing
the validity of the \emph{verification condition}~(VC)
\begin{code}
  a <= b => (cmp <=> a <= b) => v == cmp => (v<=>true)
\end{code}
The first antecedent comes from the input type of @b@, the
second from the type of @cmp@ obtained from the output of @leq@,
the third from the \emph{actual} input passed to @assert@,
and the goal comes from the input type \emph{required} by @assert@.
An SMT solver \cite{NelsonOppen} readily establishes the validity
of the above VC, thereby verifying @checkGE@.


\paragraph{First order refinements prevent modular specifications.}
Consider the function that returns the largest element of a list:
\begin{code}
  maximum         :: List Int -> Int
  maximum [x]     = x
  maximum (x:xs)  = max x (maximum xs)
    where max a b = if a < b then b else a
\end{code}
How can one write a first-order refinement type specification for
@maximum@ that will let us verify the below code?
\begin{code}
  posMax :: List Pos -> Pos
  posMax = maximum
\end{code}
%
%
Any suitable specification would have to enumerate the
situations under which @maximum@ may be invoked
breaking modularity.

\paragraph{Abstract Refinements} overcome the above modularity
problems \cite{vazou13}.
The main idea is that we can type @maximum@ by observing
that it returns \emph{one of} the elements in its input list.
Thus, if every element of the list enjoys some refinement @p@
then the output value is also guaranteed to satisfy @p@.
Concretely, we can type the function as:
\begin{code}
maximum :: forall<p::Int->Bool>. List Int<p> -> Int<p>
\end{code}
where informally, @Int<p>@ stands for @{v:Int | p v}@,
and @p@ is an \emph{uninterpreted function} in the refinement
logic~\cite{NelsonOppen}.
The signature states that for any refinement @p@ on @Int@,
the input is a list of elements satisfying @p@
and returns as output an integer satisfying @p@.
In the sequel, we will drop the explicit quantification
of abstract refinements; all free abstract refinements
will be \emph{implicitly} quantified at the top-level
(as with classical type parameters.)

\paragraph{Abstract Refinements Preserve Decidability.}
Abstract refinements do not require the use of higher-order
logics. Instead, abstractly refined signatures (like @maximum@)
can be verified by viewing the abstract refinements @p@ as
uninterpreted functions that only satisfy the axioms of
congruence, namely:
\begin{code}
  forall x y. x = y => p x <=> p y
\end{code}
As the quantifier free theory of uninterpreted functions
is decidable \cite{NelsonOppen}, abstract refinement type
checking remains decidable \cite{vazou13}.

\paragraph{Abstract Refinements are Automatically Instantiated} at call-sites,
via the abstract interpretation framework of Liquid Typing~\cite{vazou13}.
Each instantiation yields fresh refinement variables on
which subtyping constraints are generated; these constraints
are solved via abstract interpretation yielding the instantiations.
Hence, we verify @posMax@ 
by instantiating:
\begin{code}
  p |-> \ v -> 0 < v   -- at posMax
\end{code}

\subsection{Bounded Refinements}

Even with abstraction, refinement types hit various
expressiveness walls. Consider the following example
from~\cite{TerauchiPOPL13}.
@find@ takes as input a predicate @q@, a continuation
@k@ and a starting number @i@; it proceeds to compute
the smallest @Int@ (larger than @i@) that satisfies
@q@, and calls @k@ with that value.
@ex1@ passes @find@ a continuation that checks that the
``found'' value equals or exceeds @n@.
\begin{code}
  ex1 :: (Int -> Bool) -> Int -> ()
  ex1 q n = find q (checkGE n) n

  find q k i
    | q i       = k i
    | otherwise = find q k (i + 1)
\end{code}

\paragraph{Verification fails} as there is no way to specify that
@k@ is only called with arguments greater than @n@.
First, the variable @n@ is not in scope at the function
definition and so we cannot refer to it.
Second, we could try to say that @k@ is invoked with values
greater than or equal to @i@, which gets substituted with @n@
at the call-site. Alas, due to the currying order, @i@ too is
not in scope at the point where @k@'s type is defined and so
the type for @k@ cannot depend upon @i@.

\paragraph{Can Abstract Refinements Help?} Lets try to
abstract over the refinement that @i@ enjoys, and
assign @find@ the type:
\begin{code}
  (Int -> Bool) -> (Int<p> -> a) -> Int<p> -> a
\end{code}
which states that for any refinement @p@, the function takes
an input @i@ which satisfies @p@ and hence that the continuation
is also only invoked on a value which trivially enjoys @p@, namely @i@.
At the call-site in @ex1@ we can instantiate
\begin{equation}
\cc{p} \mapsto \lambda \cc{v} \rightarrow \cc{n} \leq \cc{v} \label{eq:inst:find}
\end{equation}
This instantiated refinement is satisfied by the parameter @n@, and
sufficient to verify, via function subtyping, that @checkGE n@ will
only be called with values satisfying @p@, and hence larger than @n@.

\paragraph{\cc{find} is ill-typed} as the signature requires that
at the recursive call site, the value @i+1@ \emph{also}
satisfies the abstract refinement @p@.
While this holds for the example we have in mind~(\ref{eq:inst:find}),
it does not hold \emph{for all} @p@, as required by the type of @find@!
Concretely, @{v:Int|v=i+1}@ is in general \emph{not} a subtype of
@Int<p>@, as the associated VC
%
\begin{equation}
    ... \Rightarrow \cc{p i} \Rightarrow \cc{p (i+1)} \label{eq:vc:find}
\end{equation}
%
%
is \emph{invalid} -- the type checker thus (soundly!) rejects @find@.

\paragraph{We must Bound the Quantification} of @p@ to limit
it to refinements satisfying some constraint, in this case
that @p@ is \emph{upward closed}. In the dependent setting,
where refinements may refer to program values, bounds
are naturally expressed as constraints between refinements.
%
We define a bound, @UpClosed@
which states that @p@ is a refinement that is \emph{upward closed},
\ie satisfies @forall x. p x =>  p (x+1)@,
and use it to type @find@ as:
\begin{code}
  bound UpClosed (p :: Int -> Bool)
    = \x -> p x => p (x+1)

  find :: (UpClosed p) => (Int -> Bool)
                       -> (Int<p> -> a)
                       ->  Int<p> -> a
\end{code}
This time, the checker is able to use the bound to
verify the VC~(\ref{eq:vc:find}).
We do so in a way that refinements (and thus VCs) remain quantifier
free and hence, SMT decidable~(\S~\ref{sec:overview:implementation}).

\paragraph{At the call to \cc{find}} in the body of @ex1@, we perform
the instantiation~(\ref{eq:inst:find}) which generates the
\emph{additional} VC
\hbox{(@n <=  x => n <=  x+1@)}
by plugging in the concrete refinements to the bound constraint.
The SMT solver easily checks the validity of the VC
and hence this instantiation, thereby statically
verifying @ex1@, \ie that the assertion inside
@checkGE@ cannot fail.
%

\subsection{Bounds for Higher-Order Functions}

Next, we show how bounds expand the scope of refinement typing by
letting us write precise modular specifications for various canonical
higher-order functions.

\subsubsection*{Function Composition}\label{sec:compose}

First, consider @compose@. What is a modular specification
for @compose@ that would let us verify that @ex2@ enjoys
the given specification?
\begin{code}
  compose f g x = f (g x)

  type Plus x y = {v:Int | v = x + y}
  ex2    :: n:Int -> Plus n 2
  ex2    = incr `compose` incr

  incr   :: n:Int -> Plus n 1
  incr n = n + 1
\end{code}

\paragraph{The challenge is to chain the dependencies} between the
input and output of @g@ and the input and output of @f@ to
obtain a relationship between the input and output of the
composition. We can capture the notion of chaining in a bound:
%
\begin{code}
  bound Chain p q r = \x y z ->
        q x y => p y z => r x z
\end{code}
which states that for any @x@, @y@ and @z@, if
(1) @x@ and @y@ are related by @q@, and
(2) @y@ and @z@ are related by @p@, then
(3) @x@ and @z@ are related by @r@.

We use @Chain@ to type @compose@ using three abstract
refinements @p@, @q@ and @r@, relating the arguments
and return values of @f@ and @g@ to their composed value.
(Here, @c<r x>@ abbreviates @{v:c | r x v}@.)

\begin{code}
  compose :: (Chain p q r) => (y:b -> c<p y>)
                           -> (x:a -> b<q x>)
                           -> (w:a -> c<r w>)
\end{code}

\paragraph{To verify \cc{ex2}} we instantiate, at the call to @compose@,
\begin{code}
  p, q |-> \x v -> v = x + 1
     r |-> \x v -> v = x + 2
\end{code}
The above instantiation satisfies the bound, as shown by the validity
of the VC derived from instantiating @p@, @q@, and @r@ in @Chain@:
\begin{code}
  y == x + 1 => z == y + 1 => z == x + 2
\end{code}
and hence, we can check that @ex2@ implements its specified type.

\subsubsection*{List Filtering}

Next, consider the list @filter@ function.
What type signature for @filter@ would let us check @positives@?
\begin{code}
  filter q (x:xs)
    | q x         = x : filter q xs
    | otherwise   = filter q xs
  filter _ []     = []

  positives       :: [Int] -> [Pos]
  positives       = filter isPos
    where isPos x = 0 < x
\end{code}
Such a signature would have to relate the @Bool@ returned by
@f@ with the property of the @x@ that it checks for.
Typed Racket's latent predicates~\cite{typedracket}
account for this idiom, but are a special construct
limited to @Bool@-valued ``type'' tests, and not
arbitrary invariants.
Another approach is to avoid the so-called
``Boolean Blindness'' that accompanies
@filter@ by instead using option types
and @mapMaybe@.

\paragraph{We overcome blindness using a witness} bound:
\begin{code}
  bound Witness p w = \x b -> b => w x b => p x
\end{code}
which says that @w@ \emph{witnesses} the
refinement @p@. That is, for any boolean @b@ such
that @w x b@ holds, if @b@ is @True@ then @p x@ also holds.

\paragraph{\cc{filter} can be given a type} saying that the output values
enjoy a refinement @p@ as long as the test predicate @q@ returns
a boolean witnessing @p@:
\begin{code}
  filter :: (Witness p w) => (x:a -> Bool<w x>)
                          -> List a
                          -> List a<p>
\end{code}

\paragraph{To verify \cc{positives}} we infer the following type and
instantiations for the abstract refinements @p@ and @w@ at the
call to @filter@:
\begin{code}
  isPos :: x:Int -> {v:Bool | v <=> 0 < x}
  p     |-> \v    -> 0 < v
  w     |-> \x b  -> b <=> 0 < x
\end{code}

\subsubsection*{List Folding}

Next, consider the list fold-right function. Suppose we
wish to prove the following type for @ex3@:
\begin{code}
  foldr :: (a -> b -> b) -> b -> List a -> b
  foldr op b []     = b
  foldr op b (x:xs) = x `op` foldr op b xs

  ex3 :: xs:List a -> {v:Int | v == len xs}
  ex3 = foldr (\_ -> incr) 0
\end{code}
where @len@ is a \emph{logical} or \emph{measure}
function used to represent the number of elements of
the list in the refinement logic~\cite{LiquidICFP14}:
\begin{code}
  measure len :: List a -> Nat
  len []      = 0
  len (x:xs)  = 1 + len xs
\end{code}

\paragraph{We specify induction as a bound.} Let
(1)~@inv@ be an abstract refinement relating a list @xs@
    and the result @b@ obtained by folding over it, and
(2)~@step@ be an abstract refinement relating the
    inputs @x@, @b@ and output @b'@ passed to and
    obtained from the accumulator @op@ respectively.
We state that @inv@ is closed under @step@ as:
\begin{code}
  bound Inductive inv step = \x xs b b' ->
        inv xs b => step x b b' => inv (x:xs) b'
\end{code}

\paragraph{We can give \cc{foldr} a type} that says that the
function \emph{outputs} a value that is built inductively
over the entire \emph{input} list:
\begin{code}
  foldr :: (Inductive inv step)
        => (x:a -> acc:b -> b<step x acc>)
        -> b<inv []>
        -> xs:List a
        -> b<inv xs>
\end{code}
That is, for any invariant @inv@ that is inductive
under @step@, if the initial value @b@ is @inv@-related
to the empty list, then the folded output is @inv@-related
to the input list @xs@.

\paragraph{We verify \cc{ex3}} by inferring, at the call to @foldr@
\begin{code}
  inv  |-> \xs v   -> v  == len xs
  step |-> \x b b' -> b' == b + 1
\end{code}
The SMT solver validates the VC obtained by plugging the
above into the bound.
Instantiating the signature for @foldr@ yields precisely the
output type desired for @ex3@.

Previously, \cite{vazou13} describes a way to type @foldr@
using abstract refinements that required the operator @op@
to have one extra ghost argument.
Bounds let us express induction without ghost arguments.

\subsection{Implementation}\label{sec:overview:implementation}

To implement bounded refinement typing, we must solve two
problems. Namely, how do we
(1)~\emph{check}, and
(2)~\emph{use}
functions with bounded signatures?
We solve both problems via a unifying insight inspired
by the way typeclasses are implemented in Haskell.
\begin{enumerate}
\item \emphbf{A Bound Specifies} a function type
whose inputs are unconstrained, and whose output is
some value that carries the refinement corresponding
to the bound's body.
\item \emphbf{A Bound is Implemented} by a ghost
function that returns $\true$, but is defined
in a context where the bound's constraint holds when
instantiated to the concrete refinements at the context.
\end{enumerate}

\paragraph{We elaborate bounds into ghost functions} satisfying
the bound's type.
To \emph{check} bounded functions, we need to
\emph{call} the ghost function to materialize the
bound constraint at particular values of interest.
Dually, to \emph{use} bounded functions, we need to
\emph{create} ghost functions whose outputs are
guaranteed to satisfy the bound constraint.
This elaboration reduces \emph{bounded} refinement
typing to the simpler problem
of \emph{unbounded} abstract refinement typing~\cite{vazou13}.
The formalization of our elaboration is described in
\S~\ref{sec:check}.
Next, we illustrate the elaboration by explaining how
it addresses the problems of checking and using bounded
signatures like @compose@.

\paragraph{We Translate Bounds into Function Types} called the
bound-type where the inputs are unconstrained, and the
outputs satisfy the bound's constraint.
For example, the bound @Chain@ used to type @compose@ in
\S~\ref{sec:compose}, corresponds to a function type, yielding
the translated type for @compose@:
\begin{code}
  type ChainTy p q r
    =  x:a -> y:b -> z:c
    -> {v:Bool | q x y => p y z => r x z}

  compose :: (ChainTy p q r) -> (y:b -> c<p y>)
                             -> (x:a -> b<q x>)
                             -> (w:a -> c<r w>)
\end{code}

\paragraph{To Check Bounded Functions} we view the bound constraints
as extra (ghost) function parameters (cf. type class dictionaries),
that satisfy the bound-type. Crucially, each expression where a
subtyping constraint would be generated (by plain refinement typing)
is wrapped with a ``call'' to the ghost to materialize the constraint
at values of interest. For example we elaborate @compose@ into:
\begin{code}
  compose $chain f g x =
    let t1 = g x
        t2 = f t1
        _  = $chain x t1 t2   -- materialize
    in  t2
\end{code}
In the elaborated version @$chain@ is the ghost parameter 
corresponding to the bound. As is standard \cite{LiquidPLDI08},
we perform ANF-conversion to name intermediate values, and then
wrap the function output with a call to the ghost to materialize
the bound's constraint. Consequently, the output of compose, namely
@t2@, is checked to be a subtype of the specified output type,
in an environment \emph{strengthened} with the bound's constraint
instantiated at @x@, @t1@ and @t2@. This subtyping reduces to a
quantifier free VC:
\begin{code}
      q x t1
  =>  p t1 t2
  => (q x t1 => p t1 t2 => r x t2)
  =>  v == t2 => r x v
\end{code}
whose first two antecedents are due to the types of @t1@ and @t2@
(via the output types of @g@ and @f@ respectively), and the third
comes from the call to @$chain@. 
The output value @v@ has the singleton refinement that
states it equals to @t2@, and finally the VC states that the
output value @v@ must be related to the input @x@ via @r@.
An SMT solver validates this decidable VC easily, thereby
verifying @compose@.

Our elaboration inserts materialization calls \emph{for all}
variables (of the appropriate type) that are in scope at the
given point. This could introduce upto $n^k$ calls where $k$
is the number of parameters in the bound and $n$ the number
of variables in scope. In practice (\eg in @compose@) this
number is small (\eg 1) since we limit ourselves to variables
of the appropriate types.

To preserve semantics we ensure that none of these materialization
calls can diverge, by carefully constraining the structure of
the arguments that instantiate the ghost functional parameters.

\paragraph{At Uses of Bounded Functions} our elaboration uses
the bound-type to create lambdas with appropriate parameters
that just return @true@. For example, @ex2@ is elaborated to:
\begin{code}
  ex2 = compose (\_ _ _ -> true) incr incr
\end{code}
This elaboration seems too na\"ive to be true: how do we
ensure that the function actually satisfies the bound type?

Happily, that is automatically taken care of by function subtyping.
Recalling the translated type for @compose@, the elaborated lambda
@(\_ _ _ ->  true)@ is constrained to be a subtype of @ChainTy p q r@.
In particular, given the call site instantiation
\begin{mcode}
  p $\mapsto$ \ y z -> z == y + 1
  q $\mapsto$ \ x y -> y == x + 1
  r $\mapsto$ \ x z -> z == x + 2
\end{mcode}
this subtyping constraint reduces to the quantifier-free VC:
\begin{align}
\inter{\Gamma}
  \Rightarrow \mathtt{true}
  \Rightarrow \cc{(z == y + 1)}
  & \Rightarrow \cc{(y == x + 1)}\notag\\
  & \Rightarrow \cc{(z == x + 2)} \label{vc:ex2}
\end{align}
where $\Gamma$ contains assumptions about the various binders in
scope.
The VC \ref{vc:ex2} is easily proved valid by an SMT solver, thereby
verifying the subtyping obligation defined by the bound, and hence,
that @ex2@ satisfies the given type.

%

\section{Formalism}\label{sec:check}

Next we formalize Bounded Refinement Types by defining
a core calculus \boundedcorelan and showing how it can
be reduced to \corelan, the core language of Abstract
Refinement Types~\citep{vazou13}.
We start by defining the syntax~(\S~\ref{sec:syntax-corelan})
and semantics~(\S~\ref{sec:semantics-corelan}) of \corelan
and the syntax of \boundedcorelan~(\S~\ref{sec:syntax-boundedcorelan}).
Next, we provide a translation from \boundedcorelan to
\corelan ~(\S~\ref{sec:translation}).
Then, we prove soundness by showing that our translation
is semantics preserving~(\S~\ref{sec:soundness}).
Finally, we describe how type inference remains
decidable in the presence of bounded refinements~(\S~\ref{sec:infer}).

\subsection{Syntax of \corelan}\label{sec:syntax-corelan}

\newcommand{\ra}[1]{\renewcommand{\arraystretch}{#1}}
\ra{0.9}
\begin{figure}[t!]
\centering
$$
\begin{array}{rrcl}
\emphbf{Expressions} \quad 
  & e & ::=     & x 
                  \spmid c 
                  \spmid \efunt{x}{\rtyp}{e} 
                  \spmid \eapp{e}{x}      \\
  &   &  \spmid & \elet{x}{e}{e}{\rtyp}   \\
  &   &  \spmid & \etabs{\tvar}{e}  
                  \spmid \etapp{e}{\rtyp} \\ 
  &   &  \spmid & \epabs{\rvar}{\rtyp}{e}
                  \spmid \epapp{e}{\constraint}     \\[0.03in] 
  
\emphbf{Constants} \quad
  & c 
  & ::= 
  & \true \spmid \false \spmid \ecrash \\
  && \spmid &  0 \spmid 1 \spmid -1 \spmid \dots
  \\[0.05in] 
  
\emphbf{Parametric Refinements} \quad 
& \constraint & ::= & \reft 
                        \spmid \efunt{x}{b}{\constraint}  \\[0.03in]

\emphbf{Predicates} \quad 
  & \creft & ::= & c \spmid \lnot \creft 
                   \spmid \creft = \creft 
                   \spmid 
                   \dots  \\[0.05in] 

\emphbf{Atomic Refinements} \quad 
  & \areft & ::= & \creft 
                   \spmid \rvapp{\rvar}{x} \\[0.03in] 

\emphbf{Refinements} \quad 
  & \reft & ::= & \areft 
                  \spmid \areft \wedge \reft 
                  \spmid \areft \Rightarrow \reft \\[0.03in] 

\emphbf{Basic Types} \quad 
  & b 
  & ::= & \tbint
          \spmid \tbbool
          \spmid \tvar    \\[0.03in]


\emphbf{Types} \quad 
  & \rtyp
  & ::=      & \tref{b}{\reft} \\
  & & \spmid & \trfun{x}{\rtyp}{\rtyp}{\reft} \\[0.03in]

\emphbf{Bounded Types} \quad 
  & \bt
  & ::= & \rtyp \\[0.05in]

\emphbf{Schemata} \quad 
  & \sigma
  & ::= & \bt
          \spmid \ttabs{\tvar}{\sigma}
          \spmid \tpabs{\rvar}{\rtyp}{\sigma} \\[0.03in]
\end{array}
$$
\caption{\textbf{Syntax of \corelan}}
\label{fig:syntax}
\end{figure}


\begin{figure}[t!]
$$
\begin{array}{rrcl}
\centering

\emphbf{Bounded Types} \quad 
  & \bt         & ::= & \rtyp 
                        \spmid \tconstraint{\constraint}{\bt} \\[0.03in]

\emphbf{Expressions} \quad 
  & e           & ::= & \dots 
                        \spmid \econstraint{\constraint}{e} 
                        \spmid \econstantconstraint{e}{\constraint} \\
\end{array}
$$
\caption{\textbf{Extending Syntax of \corelan to \boundedcorelan}} 
\label{fig:boundedsyntax}
\end{figure}
\ra{1.0}

We build our core language on top of \corelan, the language
of Abstract Refinement Types~\citep{vazou13}.
Figure~\ref{fig:syntax} summarizes the syntax of \corelan,
a polymorphic $\lambda$-calculus extended with abstract
refinements.

\paragraph{The Expressions} of \corelan include the usual variables $x$,
primitive constants $c$, $\lambda$-abstraction $\efunt{x}{\rtyp}{e}$,
application $\eapp{e}{e}$,
let bindings $\elet{x}{e}{e}{\rtyp}$,
type abstraction $\etabs{\alpha}{e}$,
and type application $\etapp{e}{\rtyp}$.
(We add let-binders to \corelan
from \cite{vazou13} as they can be reduced to $\lambda$-abstractions
in the usual way.)
The parameter $\rtyp$ in the type application is a \emph{refinement
type}, as described shortly.  Finally, \corelan includes refinement
abstraction $\epabs{\rvar}{\rtyp}{e}$, which introduces a refinement
variable $\rvar$ (with its type $\rtyp$), which
can appear in refinements inside $e$, and the corresponding refinement
application $\epapp{e}{\constraint}$ that substitutes an abstract refinement
with the parametric refinement $\constraint$, \ie
refinements $\reft$ closed under lambda abstractions.

\paragraph{The Primitive Constants} of \corelan include
\true, \false, @0@, @1@, @-1@, \etc. In addition, we include a
special untypable constant \ecrash that models ``going wrong''.
Primitive operations return a crash when invoked with inputs
outside their domain, \eg when @/@ is invoked with @0@ as the
divisor, or when an @assert@ is applied to \false.

\paragraph{Atomic Refinements} $\areft$ are either concrete or abstract refinements.
A \emph{concrete refinement} \creft is a boolean valued expression
(such as a constant, negation, equality, \etc)
drawn from a \emph{strict subset} of the language of expressions
which includes only terms that
(a)~neither diverge nor crash, and
(b)~can be embedded into an SMT decidable refinement logic including
the quantifier free theory of linear arithmetic and uninterpreted
functions~\cite{LiquidICFP14}.
An \emph{abstract refinement} $\rvapp{\pi}{x}$ is an application of
a refinement variable $\pi$ to a sequence of program variables.
A \emph{refinement} \reft is either a conjunction or
implication of atomic refinements.
To enable inference, we only allow implications to appear within
bounds $\constraint$ (\S~\ref{sec:infer}).

\paragraph{The Types of \corelan} written $\rtyp$ include basic types,
dependent functions and schemata quantified over type and refinement
variables $\tvar$ and $\rvar$ respectively.
A basic type is one of $\tbint$, $\tbbool$, or a type
variable $\alpha$.
A refined type $\rtyp$ is either a refined basic type $\tref{b}{\reft}$,
or a dependent function type $\trfun{x}{\rtyp}{\rtyp}{\reft}$ where
the parameter $x$ can appear in the refinements of the output type.
(We include refinements for functions, as refined type variables can be
replaced by function types. However, typechecking ensures these refinements
are trivially true.)
In \corelan bounded types $\bt$ are just a synonym for types $\rtyp$.
Finally, schemata are obtained by quantifying bounded types over type
and refinement variables.

\subsection{Semantics of \corelan}\label{sec:semantics-corelan}

\begin{figure*}[ht!]
\judgementHead{Well-Formedness}{\isWellFormed{\Gamma}{\sigma}}

$$\inference
    {\hastype{\Gamma, \vref:b}{\reft}{\tbbool}}
    {\isWellFormed{\Gamma}{\tref{b}{\reft}}}
    [\wtBase]
\qquad
\inference
    {
	\hastype{\Gamma}{\reft}{\tbbool} &&
    \isWellFormed{\Gamma}{\rtyp_x} &&
	\isWellFormed{\Gamma, x:\rtyp_x}{\rtyp}
    }
    {\isWellFormed{\Gamma}{\trfun{x}{\rtyp_x}{\rtyp}{\reft}}}
    [\wtFun]
$$

$$
\inference
  {\isWellFormed{\Gamma, \rvar:\rtyp}{\sigma}}
  {\isWellFormed{\Gamma}{\tpabs{\rvar}{\rtyp}{\sigma}}}
  [\wtPred]
\quad
\inference
    {\isWellFormed{\Gamma}{\sigma}}
    {\isWellFormed{\Gamma}{\ttabs{\alpha}{\sigma}}}
    [\wtPoly]
$$

\medskip \judgementHead{Subtyping}{\isSubType{\Gamma}{\sigma_1}{\sigma_2}}

$$
\inference
   {(\inter{\Gamma} \Rightarrow \inter{\reft_1} 
                 \Rightarrow  \inter{\reft_2})
                 \ \text{is valid}}
   {\isSubType{\Gamma}{\tref{b}{\reft_1}}{\tref{b}{\reft_2}}}
   [\tsubBase]
\qquad
\inference{
	\isSubType{\Gamma}{\rtyp_2}{\rtyp_1} &
	\isSubType{\Gamma, x_2:{\rtyp_2}}{\SUBST{\rtyp_1'}{x_1}{x_2}}{\rtyp_2'}	
   }
   {\isSubType{\Gamma}
	  {\trfun{x_1}{\rtyp_1}{\rtyp_1'}{\reft_1}}
	  {\trfun{x_2}{\rtyp_2}{\rtyp_2'}{\true}}
}[\tsubFun]
$$

$$
\begin{array}{c}
\inference
   {\isSubType{\Gamma, \rvar:\rtyp}{\sigma_1}{\sigma_2}}
   {\isSubType{\Gamma}{\tpabs{\rvar}{\rtyp}{\sigma_1}}{\tpabs{\rvar}{\rtyp}{\sigma_2}}}
   [\tsubPred]
\qquad
\inference
   {\isSubType{\Gamma}{\sigma_1}{\sigma_2}}
   {\isSubType{\Gamma}{\ttabs{\alpha}{\sigma_1}}{\ttabs{\alpha}{\sigma_2}}}
   [\tsubPoly]
\end{array}
$$

\medskip \judgementHead{Type Checking}{$\hastype{\Gamma}{e}{\sigma}$}

$$
\begin{array}{cc}
\inference
  {  \hastype{\Gamma}{e}{\sigma_2} && \isSubType{\Gamma}{\sigma_2}{\sigma_1} 
  && \isWellFormed{\Gamma}{\sigma_1}
  }
  {\hastype{\Gamma}{e}{\sigma_1}}
  [\tsub]
& 
\inference
  {\hastype{\Gamma}{e_x}{\rtyp_x} && 
   \hastype{\Gamma, x:\rtyp_x}{e}{\rtyp} && 
   \isWellFormed{\Gamma}{\rtyp}
  }
  {\hastype{\Gamma}{\elet{x}{e_x}{e}{}}{\rtyp}}
  [\tlet]
\end{array}
$$
$$
\begin{array}{ccc}

\inference
  {x: \tref{b}{r} \in \Gamma}
  {\hastype{\Gamma}{x}{\tref{b}{\vref = x}}}
  [\tbase]

&

\inference
  {x:\rtyp \in \Gamma}
  {\hastype{\Gamma}{x}{\rtyp}} 
  [\tvariable]
&
\inference
  {}
  {\hastype{\Gamma}{c}{\tc{c}}}
  [\tconst]

\\[0.2in]

\label{tapp}
\inference
   {\hastype{\Gamma}{e_1}{\tfun{x}{\rtyp_x}{\rtyp}} 
   &&  \hastype{\Gamma}{e_2}{\rtyp_x}
   }
   {\hastype{\Gamma}{\eapp{e_1}{e_2}}{\SUBST{\rtyp}{x}{e_2}}}
   [\tapp]

&

\inference
   {\hastype{\Gamma, x:\rtyp_x}{e}{\rtyp} 
    && \isWellFormed{\Gamma}{\rtyp_x}
   }
   {\hastype{\Gamma}{\efunt{x}{\rtyp_x}{e}}{\tfun{x}{\rtyp_x}{\rtyp}}}
   [\tfunction]

& 

\inference
  {\hastype{\Gamma, \alpha}{e}{\sigma}}
  {\hastype{\Gamma}{\etabs{\alpha}{e}}{\ttabs{\alpha}{\sigma}}}
  [\tgen]

\\[0.2in]

\inference
    {\hastype{\Gamma}{e}{\tpabs{\rvar}{\rtyp}{\sigma}} && 
     \hastype{\Gamma}{\efunbar{x:\rtyp_x}{\reft'}}{\rtyp}
    }
    {\hastype{\Gamma}
             {\epapp{e}{\efunbar{x:\rtyp_x}{\reft'}}}
             {\rpinst{\sigma}{\rvar}{\efunbar{x:\rtyp_x}{\reft'}}}
    }
    [\tpinst]
&

\inference
    {\hastype{\Gamma, \rvar:\rtyp}{e}{\sigma} &&
     \isWellFormed{\Gamma}{\rtyp} 
    }
    {\hastype{\Gamma}{\epabs{\rvar}{\rtyp}{e}}{\tpabs{\rvar}{\rtyp}{\sigma}}}
    [\tpgen]

&
\inference
  {\hastype{\Gamma}{e}{\ttabs{\alpha}{\sigma}} && 
   \isWellFormed{\Gamma}{\rtyp}
  }
  {\hastype{\Gamma}{\etapp{e}{\tau}}{\SUBST{\sigma}{\alpha}{\rtyp}}}
  [\tinst]

\end{array}$$
\caption{\textbf{Static Semantics: Well-formedness, Subtyping and Type Checking}}
\label{fig:rules}
\end{figure*}

Figure \ref{fig:rules} summarizes the static semantics of \corelan
as described in ~\citep{vazou13}.
Unlike~\citep{vazou13} that syntactically separates concrete ($p$)
from abstract ($\rvapp{\rvar}{x}$) refinements,
here, for simplicity, we merge both concrete and abstract refinements to
atomic refinements $\areft$.

\paragraph{A type environment} $\Gamma$ is a sequence of type bindings $x:\sigma$.
We use environments to define three kinds of judgments:

\begin{itemize}
\item{\emphbf{Well-formedness judgments} (\isWellFormed{\Gamma}{\sigma})}
state that a type schema $\sigma$ is well-formed under environment
$\Gamma$. That is, the judgment states that the refinements in $\sigma$
are boolean expressions in the environment $\Gamma$.

\item{\emphbf{Subtyping judgments} (\isSubType{\Gamma}{\sigma_1}{\sigma_2})}
state that the type schema $\sigma_1$ is a subtype of the type schema
$\sigma_2$ under environment $\Gamma$. That is, the judgment states that
when the free variables of $\sigma_1$ and $\sigma_2$ are bound to values
described by $\Gamma$, the values described by $\sigma_1$ are a subset
of those described by $\sigma_2$.

\item{\emphbf{Typing judgments} (\hastype{\Gamma}{e}{\sigma})} state that
the expression $e$ has the type schema $\sigma$ under environment $\Gamma$.
That is, the judgment states that when the free variables in $e$ are bound
to values described by $\Gamma$, the expression $e$ will evaluate to a value
described by $\sigma$.
\end{itemize}

\paragraph{The Well-formedness rules}
check that the concrete and abstract refinements are indeed $\tbbool$-valued
expressions in the appropriate environment. The key rule is \wtBase, which
checks that the refinement $\reft$ is boolean.

\paragraph{The Subtyping rules}
stipulate when the set of values described by schema $\sigma_1$ is subsumed
by (\ie contained within) the values described by $\sigma_2$.
The rules are standard except for \tsubBase, which reduces subtyping of
basic types to validity of logical implications, by translating the
refinements $r$ and the environment $\Gamma$ into logical formulas:
\begin{align*}
\inter{r}      & \defeq r &
\inter{\Gamma} & \defeq \bigwedge \{ \SUBST{r}{\vref}{x}\ |\ (x, \tref{b}{r}) \in \Gamma \}
\end{align*}
Recall that we ensure that the refinements $r$
belong to a decidable logic so that validity
checking can be performed by an off-the-self
SMT solver.

\paragraph{Type Checking Rules}
are standard except for \tpgen and \tpinst, which
pertain to abstraction and instantiation of abstract refinements.
The rule \tpgen is the same as \tfunction: we simply check the body
$e$ in the environment extended with a binding for the refinement
variable $\rvar$.
The rule \tpinst checks that the concrete refinement is of the appropriate
(unrefined) type $\tau$, and then replaces all (abstract) applications of
$\rvar$ inside $\sigma$ with the appropriate (concrete) refinement $\reft'$
with the parameters $\overline{x}$ replaced with arguments at that application.
In~\cite{vazou13} we prove the following soundness result for \corelan
which states that well-typed programs cannot crash:

\begin{lemma*}[Soundness of \corelan~\cite{vazou13}]
\label{theorem:core}
If   $\hastype{\emptyset}{e}{\sigma}$
then $e \not \goestostar{\ecrash}$.
\end{lemma*}

\subsection{Syntax of \boundedcorelan}\label{sec:syntax-boundedcorelan}

Figure~\ref{fig:boundedsyntax} shows how we obtain the syntax for
\boundedcorelan by extending the syntax of \corelan with
\emph{bounded} types.

\paragraph{The Types} of \boundedcorelan extend those of \corelan with
bounded types $\bt$, which are the types $\rtyp$ guarded by bounds
$\constraint$.

\paragraph{The Expressions} of \boundedcorelan extend those of \corelan
with \emph{abstraction} over bounds $\econstraint{\constraint}{e}$ and
\emph{application} of bounds $\econstantconstraint{e}{\constraint}$.
Intuitively, if an expression $e$ has some type $\bt$
then $\econstraint{\constraint}{e}$ has the type
$\tconstraint{\constraint}{\bt}$.
We include an explicit bound application form
$\econstantconstraint{e}{\constraint}$ to simplify
the formalization; these applied bounds are automatically
synthesized from the type of $e$, and are of the form
$\overline{\efunt{x}{\bt}{}}{\true}$.

\paragraph{Notation.}
We write
$b$,
$\tpp{b}{\rvapp{\pi}{x}}$,
$\tpref{b}{\rvapp{\pi}{x}}{\reft}$
to abbreviate
$\tref{b}{\true}$,
$\tref{b}{\rvapp{\pi}{x}\ \vref}$,
$\tref{b}{\reft \wedge \rvapp{\pi}{x}\ \vref}$
respectively.
We say a type or schema is \emph{non-refined} if all the
refinements in it are $\true$.
%
We get the \textit{shape} of a type $\rtyp$ (\ie the System-F type)
by the function $\toshape{\rtyp}$ defined:
\begin{align*}
\toshape{\tref{b}{\reft}} \defeq &\ b \\
\toshape{\trfun{x}{\rtyp_1}{\rtyp_2}{\reft}} \defeq &\ \toshape{\rtyp_1} \rightarrow \toshape{\rtyp_2}
\end{align*}

\subsection{Translation from \boundedcorelan to \corelan}
\label{sec:translation}

Next, we show how to translate a term from \boundedcorelan to
one in \corelan. We assume, without loss of generality that the
terms in \boundedcorelan are in Administrative Normal Form
(\ie all applications are to variables.)

\paragraph{Bounds Correspond To Functions} that explicitly
``witness'' the fact that the bound constraint holds at a
given set of ``input'' values.
That is we can think of each bound as a universally quantified
relationship between various (abstract) refinements; by ``calling''
the function on a set of input values $x_1,\ldots,x_n$, we get
to \emph{instantiate} the constraint for that particular set
of values.

\paragraph{Bound Environments} \benv are used by our translation
to track the set of
bound-functions (names) that are in scope at each program point.
These names are distinct from the regular program variables that
will be stored in Variable Environments \cenv.
We give bound functions distinct names so that they cannot appear
in the regular source, only in the places where calls are inserted
by our translation.
The translation ignores refinements entirely; both environments
map their names to their non-refined types.

\paragraph{The Translation is formalized} in
Figure~\ref{fig:translation} via a
relation $\txexpr{\cenv}{\benv}{e}{e'}$,
that translates the expression
$e$ in $\boundedcorelan$ into
$e'$ in $\corelan$.
Most of the rules in figure~\ref{fig:translation}
recursively translate the sub-expressions.
Types that appear inside expressions are syntactically restricted to
not contain bounds,
thus types inside expressions do not require translation.
Here we focus on the three interesting rules:

\begin{enumerate}
\item \emphbf{At bound abstractions} $\econstraint{\constraint}{e}$
 we convert the bound $\constraint$ into a bound-function
 parameter of a suitable type,
\item \emphbf{At variable binding sites} \ie $\lambda$- or let-bindings,
 we \emph{use} the bound functions to \emph{materialize} the
 bound constraints for all the variables in scope after the binding,
\item \emphbf{At bound applications} $\econstantconstraint{e}{\constraint}$
 we \emph{provide} regular functions that witness that the bound constraints hold.
\end{enumerate}



\begin{figure}[t!]
$$
\begin{array}{rrcl}
\centering
 \emphbf{Variable Environment} \quad 
   & \cenv & ::=
   & \emptyset \spmid  \EXT{\cenv}{x}{\utyp}
   \\[0.05in]
 
 \emphbf{Bound Environment} \quad 
   & \benv & ::=
   & \emptyset \spmid  \EXT{\benv}{x}{\utyp}
\end{array}
$$

\judgementHead{Translation}{\txexpr{\cenv}{\benv}{e}{e}}
$$
\inference{
}{
	\txexpr{\cenv}{\benv}{x}{x}
}[\txVar]
\qquad
\inference{
}{
	\txexpr{\cenv}{\benv}{c}{c}
}[\txCon]
$$

$$
\inference{
	\cenv' = \EXT{\cenv}{x}{\toshape{\rtyp}} && \txexpr{\cenv'}{\benv}{e}{e'} 
}{
	\txexpr{\cenv}{\benv}{\efunt{x}{\rtyp}{e}}{\efunt{x}{\rtyp}{\closure{\cenv'}{\benv}{e'}{x\colon\rtyp}}}
}[\txFun]
$$
 
$$
\inference{
	\txexpr{\cenv}{\benv}{e_x}{e_x'} && \cenv' = \EXT{\cenv}{x}{\toshape{\rtyp}} &&
	\txexpr{\cenv'}{\benv}{e}{e'}
}{
	\txexpr{\cenv}{\benv}{\elet{x}{e_x}{e}{\rtyp}}
	{\elet{x}{e_x'}{\closure{\cenv'}{\benv}{e'}}{\tau}{x\colon\rtyp}}
}[\txLet]
$$

$$
\inference{
	\txexpr{\cenv}{\benv}{e_1}{e_1'} &&
	\txexpr{\cenv}{\benv}{e_2}{e_2'}
}{
	\txexpr{\cenv}{\benv}{\eapp{e_1}{e_2}}{\eapp{e_1'}{e_2'}}
}[\txApp]
$$

$$
\inference{
	\txexpr{\cenv}{\benv}{e}{e'}
}{
	\txexpr{\cenv}{\benv}{\etabs{\alpha}{e}}{\etabs{\alpha}{e'}}
}[\txTAbs]
\
\inference{
	\txexpr{\cenv}{\benv}{e}{e'}
}{
	\txexpr{\cenv}{\benv}{\etapp{e}{\rtyp}}{\etapp{e'}{\rtyp}}
}[\txTApp]
$$

$$
\inference{
	\txexpr{\cenv}{\benv}{e}{e'}
}{
	\txexpr{\cenv}{\benv}{\epabs{\rvar}{\rtyp}{e}}{\epabs{\rvar}{\rtyp}{e'}}
}[\txPAbs]
$$
 
$$
\inference{
	\txexpr{\cenv}{\benv}{e_1}{e_2'} &&
	\txexpr{\cenv}{\benv}{e_1}{e_2'}
}{
	\txexpr{\cenv}{\benv}{\epapp{e_1}{e_2}}{\epapp{e_1'}{e_2'}}
}[\txPApp]
$$
 
$$
\inference{
	\text{fresh}\ f &&
	\txexpr{\cenv}{\benv, f\colon\toshape{\txbound{\constraint}}}{e}{e'}
}{
	\txexpr{\cenv}{\benv}{\econstraint{\constraint}{e} }{\efunt{f}{\txbound{\constraint}}{e'}}
}[\txCAbs]
$$

$$
\inference{
	\txexpr{\cenv}{\benv}{e}{e'}
}{
	\txexpr{\cenv}{\benv}{\econstantconstraint{e}{\constraint}}{\eapp{e'}{\ctofun{\constraint}}}
}[\txCApp]
$$
\caption{\textbf{Translation Rules from \boundedcorelan to  \letcorelan}}
\label{fig:translation}
\end{figure}

\paragraph{1.  Rule~\txCAbs} translates bound abstractions
$\econstraint{\phi}{e}$ into a plain $\lambda$-abstraction.
In the translated expression $\efunt{f}{\txbound{\constraint}}{e'}$
the bound becomes a function named $f$ with type
$\txbound{\constraint}$ defined:
\begin{align*}
\txbound{\efunt{x}{b}{\constraint}} \defeq & \tfun{x}{b}{\txbound{\constraint}}\\
\txbound{r} \defeq & \tref{\tbbool}{r}
\end{align*}
That is, $\txbound{\constraint}$ is a function type whose
final output carries the refinement corresponding to
the constraint in $\constraint$.
Note that the translation generates a fresh name $f$ for
the bound function (ensuring that it cannot be used in
the regular code) and saves it in the bound environment
$\benv$ to let us materialize the bound constraint when
translating the body $e$ of the abstraction.



\paragraph{2. Rules~\txFun and~\txLet} materialize bound
constraints at variable binding sites ($\lambda$-abstractions
and let-bindings respectively.)
%
%
If we view the bounds as universally quantified constraints
over the (abstract) refinements, then our translation exhaustively
and eagerly \emph{instantiates} the constraints at each point that
a new binder is introduced into the variable environment, over all
the possible candidate sets of variables in scope at that point.
The instantiation is performed by $\closure{\cenv}{\benv}{e}{x:\rtyp}$
%
$$\begin{array}{rcl}
\closure{\cenv}{\benv}{e}{\tbind{x}{\utyp}}
  & \defeq & {\wraplet{e}{\cands{\cenv}{\benv}}} \\[0.05in]
\wraplet{e}{\set{e_1,\ldots,e_n}}
  & \defeq & {\elett{t_1}{e_1}{\ldots} \elett{t_n}{e_n}{e}} \\
  &        & {\mbox{(where $t_i$ are fresh \tbbool binders)}} \\[0.05in]
\cands{\cenv}{\benv}
  & \defeq & \{ \  f\ \overline{x}  \ | \ \tbind{f}{\utyp} \leftarrow \benv
                  , \ \overline{\tbind{x}{\_}} \leftarrow \cenv \\
  &        & \qquad \ \ , \ \EXT{\cenv}{f}{\utyp} \vdash_B \tbind{ f\ \overline{x}}{\tbbool} \} \
\end{array}$$
The function takes the environments
$\cenv$ and $\benv$, an expression $e$ and a variable $x$ of type
$\rtyp$ and uses let-bindings to materialize all the bound
functions in $\benv$ that accept the variable $x$.
Here, $\cenv \vdash_B \tbind{e}{\utyp}$ is the standard typing
derivation judgment for the non-refined System F and so
we elide it for brevity.

\paragraph{3. Rule~\txCApp} translates bound applications
$\econstantconstraint{e}{\constraint}$ into plain $\lambda$
abstractions that witness that the bound constraints
hold.
That is, as within $e$, bounds are translated to a bound
function (parameter) of type $\txbound{\constraint}$, we
translate $\constraint$ into a $\lambda$-term that, via
subtyping must have the required type $\txbound{\constraint}$.
We construct such a function via $\ctofun{\constraint}$
that depends only on the \emph{shape} of the bound,
\ie the non-refined types of its parameters (and not
the actual constraint itself).
\begin{align*}
\ctofun{\reft} \defeq & \true \\
\ctofun{\efunt{x}{b}{\constraint}} \defeq &  \efunt{x}{b}{\ctofun{\constraint}}
\end{align*}
This seems odd: it is simply a constant function, how
can it possibly serve as a bound? The answer is that
subtyping in the translated \corelan term will verify
that in the context in which the above constant function
is created, the singleton $\true$ will indeed carry
the refinement corresponding to the bound constraint,
making this synthesized constant function a valid
realization of the bound function.
Recall that in the example @ex2@ of the overview (\S~\ref{sec:overview:implementation})
the subtyping constraint that decides is the constant $\true$
is a valid bound reduces to the equation \ref{vc:ex2}
that is a tautology.

\subsection{Soundness}\label{sec:soundness}

\paragraph{The Small-Step Operational Semantics} of \boundedcorelan
are defined by extending a similar semantics for \corelan
which is a standard call-by-value calculus where abstract
refinements are boolean valued functions~\cite{vazou13}.
Let $\stepcore$ denote the transition relation defining
the operational semantics of \corelan and \tclos{\stepcore}
denote the reflexive transitive closure of $\stepcore$.
We thus obtain the transition relation $\stepboundedcore$:
\begin{align*}
\econstantconstraint{(\econstraint{\constraint}{e})}{\constraint} &\boundedgoesto{e} &
e  & \stepboundedcore e', \text{if}\ {e \stepcore e'}
\end{align*}
%
Let $\boundedgoestostar{}$ denote the reflexive transitive
closure of $\stepboundedcore$.

\paragraph{The Translation is Semantics Preserving} in the sense that
if a source term $e$ of $\boundedcorelan$ reduces to a constant
then the translated variant of $e'$ also reduces to the same
constant:

\begin{lemma*}\ifextended{[Semantics Preservation]}{}
\label{theorem:operational}
If $\txexpr{\emptyset}{\emptyset}{e}{e'}$ and
   $e \boundedgoestostar{c}$
then $e' \goestostar{c}$.
\end{lemma*}

\ifextended{
\begin{proof}
By assumption, there exists a sequence
$e \equiv e_1 \boundedgoesto{e_2} \boundedgoesto{} \dots
\boundedgoesto{e_n\equiv c} $.
Let $i$ be the largest index in which rule \rtobound was applied.
Then, for some $\phi$ and $\phi'$,
$e_i$ contains a sub-expression of the form
$\econstantconstraint{(\econstraint{\phi}{e_i^0})}{\phi'}$.
Let $e_i^1$ be the expression we get if we replace
$\econstantconstraint{(\econstraint{\phi}{e_i^0})}{\phi'}$
with $e_i^0$ in $e_i$.
By the way we choose $i$, there exist a sequence
$e_i^1 \goestostar{c}$.

Let $e_i^2$
be the expression we get if we replace
$\econstantconstraint{(\econstraint{\phi}{e_i^0})}{\phi'}$
with $\eapp{(\efunt{f}{\txbound{\phi}}{e_i^0})}{(\ctofun{\phi'})}$ in $e_i$.
Then, since $f$ does not appear in $e_i^0$,
$e_i^2 \goestostar{c}$.
Finally,
let $g \defeq \ctofun{\phi'}$, then
by the definition of
$\ctofun{\cdot}$
we have that  $\forall e_1 \dots e_n$
if
there exists a type $\tau$ such that
$\emptyset \vdash g \ e_1 \dots e_n : \tau $,
then $g \ e_1 \dots e_n \goestostar{true}$.
Thus, for any expression,
if $e \goestostar{c}$, then $\elett{t}{f \ e_1 \dots e_n}{e}\goestostar{c}$

From the above, by the way we choose $i$ we have that
there exists a sequence
$\txex{e_i} \hookrightarrow \dots \hookrightarrow {c}$.

Since $n$ is finite, we iteratively apply the above procedure to
$e \equiv e_1\boundedgoesto{} \dots \boundedgoesto{} \txex{e_i}\hookrightarrow \dots \hookrightarrow {c}$.
until we get the sequence $ {\txex{e}}\goestostar{c}$.
\end{proof}}{}

\paragraph{The Soundness of \boundedcorelan} follows by combining
the above Semantics Preservation
Lemma 
with the soundness of \corelan.

\paragraph{To Typecheck a \boundedcorelan program} $e$ we translate it
into a \corelan program $e'$ and then type check $e'$; if the latter
check is safe, then we are guaranteed that the source term $e$ will
not crash:

\begin{theorem*}[Soundness]
\label{theorem:bounded}
If $\txexpr{\emptyset}{\emptyset}{e}{e'}$ and
   $\hastype{\emptyset}{e'}{\sigma}$
then $e \not \boundedgoestostar{\ecrash}$.
\end{theorem*}

\subsection{Inference}\label{sec:infer}

A critical feature of bounded refinements is that we can
automatically synthesize instantiations of the abstract
refinements by:
(1)~generating templates corresponding to the unknown types
    where fresh variables $\kvar$ denote the unknown refinements
    that an abstract refinement parameter $\rvar$ is instantiated
    with,
(2)~generating subtyping constraints over the resulting templates,
    and
(3)~solving the constraints via abstract interpretation.

\paragraph{Inference Requires Monotonic Constraints.}
Abstract interpretation only works if the constraints
are \emph{monotonic}~\citep{cousotcousot77}, which in this case
means that the $\kvar$ variables, and correspondingly,
the abstract refinements $\rvar$ must only appear in
\emph{positive} positions within refinements (\ie not
under logical negations).
The syntax of refinements shown in Figure~\ref{fig:syntax}
violates this requirement via refinements of the
form $\rvapp{\rvar}{x} \Rightarrow \reft$.
%

\paragraph{We restrict implications to bounds} \ie prohibit
them from appearing elsewhere in type specifications.
Consequently, the implications only appear in the
\emph{output} type of the (first order) ``ghost''
functions that bounds are translated to.
The resulting subtyping constraints only have
\emph{implications inside super-types}, \ie as:
$$
\isSubType{\Gamma}{\stref{b}{\areft}}{\stref{b}{\areft_1 \Rightarrow \dots \Rightarrow \areft_{n} \Rightarrow\areft_q}}
$$
By taking into account the semantics of subtyping, we can
push the antecedents into the environment, \ie transform
the above into an equivalent constraint in the form:
$$
\isSubType{
\EXTT{
 \EXTT{\Gamma}{x_1}{\sxref{b_1}{\areft_1'}{x_1}},\dots
}{x_n}{\sxref{b_n}{\areft_n'}{x_n}}
}
{\stref{b}{\areft'}}
{\stref{b}{\areft_q'}}
$$
where all the abstract refinements variables $\rvar$
(and hence instance variables $\kvar$) appear positively,
ensuring that the constraints are monotonic, hence permitting
inference via Liquid Typing~\citep{LiquidPLDI08}.

\section{A Refined Relational Database}\label{sec:database}

Next, we use bounded refinements to develop a library
for relational algebra, which we use to enable generic,
type safe database queries.
A relational database stores data in \emph{tables},
that are a collection of \emph{rows}, which in turn 
are \emph{records} that represent a unit of data 
stored in the table.
The tables's \textit{schema} describes the types of 
the values in each row of the table.
For example, the table in Figure~\ref{fig:moviedb} organizes 
information about movies, and has the schema:
\begin{code}
 Title:String, Dir:String, Year:Int, Star:Double
\end{code}

\begin{figure}[t]
$$
\begin{tabular}{| l | l| r | r |}
  \hline
  \textbf{Title} & \textbf{Director} & \textbf{Year} & \textbf{Star} \\
  \hline  
  ``Birdman'' & ``I\~{n}\'{a}rritu''   & 2014 & 8.1\\
  ``Persepolis''  & ``Paronnaud'' & 2007 & 8.0 \\ 
  \hline
\end{tabular}
$$
\caption{\label{fig:moviedb} Example Table of Movies}
\end{figure}

First, we show how to write type safe extensible 
records  that represent rows, and use them to 
implement database tables~(\S~\ref{subsec:records}). 
Next, we show how bounds let us specify type 
safe relational operations and how they may be 
used to write safe database queries~(\S~\ref{subsec:relational}).

\subsection{Rows and Tables}\label{subsec:records}

We represent the rows of a database with dictionaries, 
which are maps from a set of keys to values.
In the sequel, each key corresponds to a column, and 
the mapped value corresponds to a valuation of the column 
in a particular row.

\paragraph{A dictionary} @Dict <r> k v@ maps a key @x@ of 
type @k@ to a value of type @v@ that satisfies the property @r x@
%
\begin{code}
  type Range k v = k -> v -> Bool
   
  data Dict k v <r :: Range k v> = D {
      dkeys :: [k]
    , dfun  :: x:{k | x Set_mem elts dkeys} -> v<r x>
    }
\end{code}      
Each dictionary @d@ has a domain @dkeys@ 
\ie the list of keys for which @d@ is defined 
and a function @dfun@ that is defined only on
elements @x@ of the domain @dkeys@.
For each such element @x@, @dfun@ returns a value that satisfies the
property @r x@.

\paragraph{Propositions about the theory of sets} can be decided
efficiently by modern SMT solvers. Hence we use such 
propositions within refinements~\citep{realworldliquid}.
The measures (logical functions) @elts@ and @keys@ 
specify the set of keys in a list and a dictionary 
respectively:
\begin{code}
  elts        :: [a] -> Set a
  elts ([])   = Set_emp
  elts (x:xs) = {x} Set_cup elts xs

  keys        :: Dict k v -> Set k
  keys d      = elts (dkeys d) 
\end{code}

\paragraph{Domain and Range of dictionaries.}
In order to precisely define the domain (\eg columns) and range (\eg values)
of a dictionary (\eg row), we define the following aliases:
%
\begin{code}
  type RD k v <dom :: Dom k v, rng :: Range k v>
    = {v:Dict <rng> k v | dom v}

  type Dom k v = Dict k v -> Bool 
\end{code}
We may instantiate @dom@ and @rng@ with predicates that precisely describe
the values contained with the dictionary.
For example,
\begin{code}
  RD < \d -> keys d == {"x"}
     , \k v-> 0 < v         > String Int
\end{code}
%
%
describes dictionaries with a single field @"x"@ 
whose value (as determined by @dfun@) is stricly 
greater than 0.
We will define schemas by appropriately 
instantiating the abstract refinements 
@dom@ and @rng@.

\paragraph{An empty dictionary} has an empty domain 
and a function that will never be called:
\begin{code}
  empty   :: RD <emptyRD, rFalse> k v
  empty   = D [] (\x -> error "calling empty")

  emptyRD = \d -> keys d == Set_emp
  rFalse  = \k v -> false
\end{code}
 
\paragraph{We define singleton maps} as dependent pairs 
@x := y@ which denote the mapping from @x@ to @y@:
\begin{code}
  data P k v <r :: Range k v> 
    = (:=) {pk :: k, pv :: v<r pk>}
\end{code}
Thus, @key := val@ has type \hbox{@P<r> k v@} only if 
@r key val@.

\paragraph{A dictionary may be extended} with a singleton
binding (which maps the new key to its new value). 
\begin{code}
  (+=)   :: bind:P<r> k v 
         -> dict:RD<pTrue, r> k v 
         -> RD <addKey (pk bind) dict, r> k v
 
  (k := v) += (D ks f) 
         = D (k:ks) 
             (\i -> if i == k then v else f i)
  
  addKey = \k d d' -> keys d' == {k} Set_cup keys d
  pTrue  = \_ -> true
\end{code}
Thus, @(k := v)  += d@ evaluates to 
a dictionary @d'@ that extends @d@ 
with the mapping from @k@ to @v@.
The type of @(+=)@ constrains the new binding @bind@, 
the old dictionary @dict@ and the returned value to have 
the same range invariant @r@.
The return type states that the output dictionary's 
domain is that of the domain of @dict@ extended by 
the new key @(pk bind)@.

\paragraph{To model a row in a table} \ie a schema, 
we define the unrefined (\Haskell) type @Schema@, 
which is a dictionary mapping @String@s, \ie the 
names of the fields of the row, to elements of 
some universe @Univ@ containing @Int@, @String@ 
and @Double@.
(A closed universe is not a practical restriction; 
most databases support a fixed set of types.)
\begin{code}
  data Univ   = I Int | S String | D Double

  type Schema = RD String Univ
\end{code}

\paragraph{We refine Schema} with concrete instantiations
for @dom@ and @rng@, in order to recover precise 
specifications for a particular database. For example, 
@MovieSchema@ is a refined @Schema@ that describes the 
rows of the Movie table in Figure~\ref{fig:moviedb}:
\begin{code}
type MovieSchema = RD <md, mr> String Univ

  md = \d -> 
      keys d={"year","star","dir","title"}
  mr = \k v -> 
      (k = "year"  => isI v && 1888 < toI v)
   && (k = "star"  => isD v && 0 <= toD v <= 10)
   && (k = "dir"   => isS v)
   && (k = "title" => isS v)

  isI (I _)   = True 
  isI _       = False 

  toI       :: {v: Univ | isI v} -> Int
  toI (I n) = n
...
\end{code}
The predicate @md@ describes the \emph{domain} of the movie schema,
restricting the keys to exactly @"year"@, @"star"@, @"dir"@, and @"title"@.
The range predicate @mr@ describes the types of the values in the schema:
a dictionary of type @MovieSchema@ must map 
@"year"@ to an @Int@,
@"star"@ to a @Double@, 
and @"dir"@ and @"title"@ to @String@s.
The range predicate may be used to impose additional constraints on the values
stored in the dictionary.
For instance, @mr@ restricts the year to be not only an integer but
also greater than @1888@.
%

\paragraph{We populate the Movie Schema} by extending the
empty dictionary with the appropriate pairs of fields and 
values. For example, here are the rows from the table
in Figure~\ref{fig:moviedb}
\begin{code}
  movie1, movie2 :: MovieSchema
  movie1 = ("title" := S "Persepolis")
        += ("dir"   := S "Paronnaud") 
        += ("star"  := D 8) 
        += ("year"  := I 2007) 
        += empty

  movie2 = ("title" := S "Birdman") 
        += ("star"  := D 8.1) 
        += ("dir"   := S "Inarritu")
        += ("year"  := I 2014) 
        += empty
\end{code}
Typing @movie1@ (and @movie2@) as @MovieSchema@
boils down to proving:
That @keys movie1 = {"year", "star", "dir", "title"}@;
and that each key is mapped to an appropriate value 
as determined by @mr@.
For example, declaring @movie1@'s year to be @I 1888@
or even misspelling @"dir"@ as @"Dir"@
will cause the @movie1@ to become ill-typed.
As the (sub)typing relation depends on logical 
implication (unlike in @HList@ based approaches 
\cite{heterogeneous}) key ordering \emph{does not} 
affect type-checking;
in @movie1@ the star field is added before the 
director, while @movie2@ follows the opposite order.

\paragraph{Database Tables} are collections of rows, 
\ie collections of refined dictionaries.
We define a type alias @RT s@ (Refined Table) for 
the list of refined dictionaries from the field 
type @String@ to the @Univ@erse.
\begin{code}
  type RT (s :: {dom::TDom, rng::TRange}) 
    = [RD <s.dom, s.rng> String Univ]

  type TDom   = Dom   String Univ
  type TRange = Range String Univ
\end{code}
For brevity we pack both the domain and the range 
refinements into a record @s@ that describes the 
schema refinement; \ie each row dictionary has 
domain @s.dom@ and range @s.rng@.

For example, the table from Figure~\ref{fig:moviedb}
can be represented as a type @MoviesTable@ which 
is an @RT@ refined with the domain and range @md@ 
and @mr@ described earlier (\S~\ref{subsec:records}):
\begin{code}
  type MoviesTable = RT {dom = md, rng = mr}
   
  movies :: MoviesTable 
  movies = [movie1, movie2]
\end{code}

\subsection{Relational Algebra}\label{subsec:relational}

Next, we describe the types of the relational algebra operators
which can be used to manipulate refined rows and tables.
For space reasons, we show the \emph{types} of the basic 
relational operators; their (verified) implementations 
can be found online~\cite{liquidhaskellgithub}.
\begin{code}
  union   :: RT s -> RT s -> RT s
  diff    :: RT s -> RT s -> RT s
  select  :: (RD s -> Bool) -> RT s -> RT s
  project :: ks:[String] -> RTSubEqFlds ks s 
          -> RTEqFlds ks s
  product :: ( Disjoint s1 s2, Union s1 s2 s
             , Range s1 s, Range s2 s) 
          => RT s1 -> RT s2 -> RT s
\end{code}

\paragraph{\texttt{union} and \texttt{diff}} compute the union 
and difference, respectively of the (rows of) two tables.
The types of @union@ and @diff@ state that the 
operators work on tables with the same schema 
@s@ and return a table with the same schema.

\paragraph{\texttt{select}} takes a predicate @p@
and a table @t@ and filters the rows of @t@ 
to those which that satisfy @p@.
The type of @select@ ensures that @p@ will 
not reference columns (fields) that are
not mapped in @t@, as the predicate @p@
is constrained to require a dictionary 
with schema @s@.

\paragraph{\texttt{project}} takes
a list of @String@ fields @ks@ 
and a table @t@ and projects 
exactly the fields @ks@ at 
each row of @t@.
@project@'s type states that for 
any schema @s@, the input table 
has type @RTSubEqFlds ks s@ 
\ie its domain should have at 
least the fields @ks@ and the 
result table has type @RTEqFlds ks s@, 
\ie its domain has exactly the elements @ks@. 
\begin{code}
  type RTSubEqFlds ks s
    = RT s{dom = \z -> elts ks Set_sub  keys z}

  type RTEqFlds ks s
    = RT s{dom = \z -> elts ks == keys z}
\end{code}
The range of the argument and the result tables 
is the same and equal to @s.rng@.

\paragraph{\texttt{product}} takes two tables 
as input and returns their (Cartesian) 
product.
It takes two Refined Tables with schemata 
@s1@ and @s2@ and returns a Refined Table 
with schema @s@. Intuitively, the output
schema is the ``concatenation'' of the input
schema; we formalize this notion using bounds:
(1)~@Disjoint s1 s2@ says the domains of 
    @s1@ and @s2@ should be disjoint,
(2)~@Union s1 s2 s@ says the domain of @s@ 
    is the union of the domains of @s1@ and @s2@, 
(3)~@Range s1 s@ (\resp @Range s2 s2@) says 
    the range of @s1@ should imply the result 
    range @s@; together the two imply the output
    schema @s@ preserves the type of each key in 
    @s1@ or @s2@.
\begin{code}
  bound Disjoint s1 s2 = \x y -> 
    s1.dom x => s2.dom y => keys x Set_cap keys y == Set_emp
   
  bound Union s1 s2 s = \x y v -> 
    s1.dom x => s2.dom y 
             => keys v == keys x Set_cup keys y 
             => s.dom v

  bound Range si s = \x k v -> 
    si.dom x => k Set_mem keys x => si.rng k v 
             => s.rng k v 
\end{code}


Thus, bounded refinements  enable the precise 
typing of  relational algebra operations.
They let us describe precisely when union, 
intersection, selection, projection and products 
can be computed, and let us determine, at compile
time the exact ``shape'' of the resulting tables.


\paragraph{We can query Databases} by writing functions 
that use the relational algebra combinators. 
For example, here is a query that returns the 
``good'' titles -- with more than 8 stars -- 
from the @movies@ table
\footnote{More example queries can be found online~\cite{liquidhaskellgithub}}
\begin{code}
  good_titles = project ["title"] $ select (\d ->
                  toDouble (dfun d $ "star") > 8
                ) movies
\end{code}
%
%

Finally, note that our entire library -- including 
records, tables, and relational combinators -- is 
built using vanilla Haskell \ie without \emph{any} 
type level computation. 
All schema reasoning happens at the granularity of 
the logical refinements. That is if the refinements
are erased from the source, we still have a well-typed
Haskell program but of course, lose the safety 
guarantees about operations (\eg ``dynamic'' key lookup) 
never failing at run-time.


\section{A Refined IO Monad}\label{sec:state}

Next, we illustrate the expressiveness of Bounded Refinements by 
showing how they enable the specification and verification of 
stateful computations. We show how to 
(1)~implement a refined \emph{state transformer} 
    (\RIO) monad, where the transformer is indexed by refinements 
    corresponding to \emph{pre}- and \emph{post}-conditions 
    on the state~(\S~\ref{subsec:state:definition}),
(2)~extend \RIO with a set of combinators for 
    \emph{imperative} programming, \ie whose types 
    precisely encode Floyd-Hoare style program 
    logics~(\S~\ref{subsec:state:examples}) and
(3)~use the \RIO monad to write \emph{safe scripts}
    where the type system precisely tracks capabilities
    and statically ensures that functions only access 
    specific resources~(\S~\ref{subsec:state:files}).


\subsection{The \RIO Monad}
\label{subsec:state:definition}

\paragraph{The \RIO data type} describes stateful computations.
Intuitively, a value of type @RIO a@ denotes a computation 
that, when evaluated in an input @World@ produces a value 
of type @a@ (or diverges) and a potentially transformed 
output @World@. We implement @RIO a@ as an abstractly
refined type (as described in ~\citep{vazou13})
%
%
\begin{code}
  type Pre    = World -> Bool 
  type Post a = World -> a -> World -> Bool 

  data RIO a <p :: Pre, q :: Post a> = RIO { 
    runState :: w:World<p> -> (x:a, World<q w x>) 
  }
\end{code}
That is, @RIO a@ is a function @World-> (a, World)@, where
@World@ is a primitive type that represents the state of 
the machine \ie the console, file system, \etc
This indexing notion is directly inspired by the method 
of~\citep{Filliatre98} (also used in \cite{ynot}).


\paragraph{Our Post-conditions are Two-State Predicates}
that relate the input- and output- world (as in~\cite{ynot}). 
Classical Floyd-Hoare logic, in contrast,
uses assertions which are single-state 
predicates.
We use two-states to smoothly account for 
specifications for stateful procedures. 
This increased expressiveness makes the 
types slightly more complex than a direct
one-state encoding which is, of course 
also possible with bounded refinements.

\paragraph{An \texttt{RIO} computation is parameterized} by two 
abstract refinements:
\begin{inparaenum}[(1)]
\item @p :: Pre@, which is a predicate over the \emph{input} 
   world, \ie the input world @w@ satisfies the refinement 
   @p w@; and
\item @q :: Post a@, which is a predicate relating the 
   \emph{output} world with the input world and the 
   value returned by the computation, \ie the output 
   world @w'@ satisfies the refinement @q w x w'@ where 
   @x@ is the value returned by the computation.
\end{inparaenum}
Next, to use @RIO@ as a monad, we define @bind@ and 
@return@ functions for it, that satisfy the monad laws.
  
\paragraph{The \return operator} yields a pair of the 
supplied value @z@ and the input world unchanged:
\begin{code}
  return   :: z:a -> RIO <p, ret z> a
  return z = RIO $ \w -> (z, w)

  ret z    = \w x w' -> w' == w && x == z
\end{code}
The type of \return states that for any 
precondition @p@ and any supplied value 
@z@ of type @a@, the expression @return z@ 
is an \RIO computation with precondition
@p@ and a post-condition @ret z@.
The postcondition states that: 
(1)~the output @World@ is the same as the input, and 
(2)~the result equals to the supplied value @z@.
Note that as a consequence of the equality of the two worlds
and congruence, the output world @w'@ trivially satisfies @p w'@.
%
 
\paragraph{The \bind Operator} is defined in the usual way.
However, to type it precisely, we require bounded refinements.
\begin{code}
  (>>=) :: (Ret q1 r, Seq r q1 p2, Trans q1 q2 q)
        => m:RIO <p, q1> a
        -> k:(x:a<r> -> RIO <p2 x, q2 x> b)
        -> RIO <p, q> b 

  (RIO g) >>= f = RIO $ \x -> 
    case g x of { (y, s) -> runState (f y) s } 
\end{code}
The bounds capture various sequencing requirements 
(c.f. the Floyd-Hoare rules of consequence).
First, the output of the first action @m@, 
satisfies the refinement required by the 
continuation @k@;
\begin{code}
  bound Ret q1 r = \w x w' -> q1 w x w' => r x 
\end{code}
Second, the computations may be sequenced,
\ie the postcondition of the first action 
@m@ implies the precondition of the 
continuation @k@ (which may be dependent 
upon the supplied value @x@):
\begin{code}
  bound Seq q1 p2 = \w x w' -> 
        q1 w x w' => p2 x w'
\end{code}%
Third, the transitive composition of the two 
computations, implies the final postcondition:
\begin{code}
  bound Trans q1 q2 q = \w x w' y w'' -> 
        q1 w x w' => q2 x w' y w'' => q w y w''
\end{code}
  
Both type signatures would be impossible 
to use if the programmer had to manually 
instantiate the abstract refinements 
(\ie pre- and post-conditions.) 
Fortunately, Liquid Type inference 
generates the instantiations making it practical
to use \toolname to verify stateful computations
written using @do@-notation.

\subsection{Floyd-Hoare Logic in the \RIO Monad}
\label{subsec:state:examples}

Next, we use bounded refinements to derive an
encoding of Floyd-Hoare logic, by showing how to 
read and write (mutable) variables and
typing higher order 
@ifM@ and @whileM@ combinators.

\paragraph{We Encode Mutable Variables} as fields of 
the @World@ type. For example, we might encode
a global counter as a field:
\begin{code}
  data World = { ... , ctr :: Int, ... }
\end{code}
We encode mutable variables in the refinement logic
using McCarthy's @select@ and @update@ operators 
for finite maps and the associated axiom:
\begin{code}
  select :: Map k v -> k -> v
  update :: Map k v -> k -> v -> Map k v

  forall m, k1, k2, v.
       select (update m k1 v) k2
    == (if k1 == k2 then v else select m k2 v)
\end{code}
The quantifier free theory of @select@ and @update@ is decidable
and implemented in modern SMT solvers~\cite{SMTLIB2}.

%
\paragraph{We Read and Write Mutable Variables} via 
suitable ``get'' and ``set'' actions. For example,
we can read and write @ctr@ via:
\begin{code}
  getCtr   :: RIO <pTrue, rdCtr> Int
  getCtr   = RIO $ \w -> (ctr w, w)
    
  setCtr   :: Int -> RIO <pTrue, wrCtr n> ()
  setCtr n = RIO $ \w -> ((), w { ctr = n })
\end{code}
Here, the refinements are defined as:
\begin{code}
  pTrue = \w -> True
  rdCtr = \w x w' -> w' == w && x == select w ctr
  wrCtr n = \w _ w' -> w' == update w ctr n 
\end{code}
Hence, the post-condition of @getCtr@ states 
that it returns the current value of @ctr@, 
encoded in the refinement logic with McCarthy's 
@select@ operator while leaving the world unchanged.
The post-condition of @setCtr@ states that @World@
is updated at the address corresponding to @ctr@,
encoded via McCarthy's @update@ operator. 

\paragraph{The \texttt{ifM} combinator} 
takes as input a @cond@ action that returns a @Bool@ and,
depending upon the result, executes either
the @then@ or @else@ actions. We type it as:
%
\begin{code}
  bound Pure g = \w x v  ->(g w x v => v == w)
  bound Then g p1 = \w v -> (g w True  v => p1 v)
  bound Else g p2 = \w v -> (g w False v => p2 v)

  ifM :: (Pure g, Then g p1, Else g p2)
      => RIO <p , g> Bool       -- cond
      -> RIO <p1, q> a          -- then
      -> RIO <p2, q> a          -- else
      -> RIO <p , q> a
\end{code}
The abstract refinements and bounds 
correspond exactly to the hypotheses in the 
Floyd-Hoare rule for the @if@ statement.
The bound @Pure g@ states that the @cond@ 
action may access but does not \emph{modify} 
the @World@, \ie the output is the same 
as the input @World@. (In classical Floyd-Hoare 
formulations this is done by syntactically 
separating terms into pure \emph{expressions} 
and side effecting \emph{statements}).
The bound @Then g p1@ and @Else g p2@ respectively
state that the preconditions of the @then@ and @else@
actions are established when the @cond@ returns @True@
and @False@ respectively. 



\paragraph{We can use \texttt{ifM}} to implement a stateful 
computation that performs a division, after checking 
the divisor is non-zero.
We specify that @div@ should not be called with a zero divisor. 
Then, \toolname verifies that @div@ is called safely:
\begin{code}
  div :: Int -> {v:Int | v /= 0} -> Int

  ifTest :: RIO Int
  ifTest = ifM nonZero divX (return 10)
    where nonZero = getCtr >>= return . (/= 0)
          divX    = getCtr >>= return . (div 42)
\end{code}
Verification succeeds as the post-condition of @nonZero@
is instantiated to 
\hbox{@\_ b w -> b <=> select w ctr /= 0@}
and the pre-condition of @divX@'s is instantiated to
\hbox{@\w -> select w ctr /= 0@}, which suffices to 
prove that @div@ is only called with non-zero values.

\paragraph{The \texttt{whileM} combinator} 
formalizes loops as @RIO@ computations:
\begin{code}
  whileM :: (OneState q, Inv p g b, Exit p g q)  
         => RIO <p, g> Bool      -- cond 
         -> RIO <pTrue, b> ()    -- body
         -> RIO <p, q> ()
\end{code}
As with @ifM@, the hypotheses of the Floyd-Hoare derivation rule
become bounds for the signature.
Given a @cond@ition with pre-condition @p@ and 
post-condition @g@ and @body@ with a true 
precondition and post-condition @b@, the computation 
@whileM cond body@ has precondition @p@ and post-condition 
@q@ as long as the bounds (corresponding to the Hypotheses
in the Floyd-Hoare derivation rule) hold.
First, @p@ should be a loop invariant; \ie when 
the @cond@ition returns @True@ the post-condition 
of the body @b@ must imply the @p@:
\begin{code}
  bound Inv p g b = \w w' w'' ->
      p w => g w True w' => b w' () w'' => p w'' 
\end{code}
Second, when the @cond@ition returns @False@ the invariant @p@
should imply the loop's post-condition @q@:
\begin{code}
  bound Exit p g q = \w w' ->
        p w => g w False w' => q w () w'
\end{code}
Third, to avoid having to transitively connect the guard and the body,
we require that the loop post-condition be a one-state predicate,
independent of the input world (as in Floyd-Hoare logic):
\begin{code}
  bound OneState q = \w w' w'' ->
        q w () w'' => q w' () w''
\end{code}

\paragraph{We can use \texttt{whileM}} to implement a loop that repeatedly
decrements a counter while it is positive, and to then verify that
if it was initially non-negative, then
at the end the counter is equal to @0@.
\begin{code}
  whileTest   :: RIO <posCtr, zeroCtr> ()
  whileTest = whileM gtZeroX decr
    where gtZeroX = getCtr >>= return . (> 0)
  
  posCtr  = \w -> 0 <= select w ctr
  zeroCtr = \_ _ w' -> 0 == select w ctr 
\end{code}
Where the decrement is implemented by @decr@ with type:
\begin{code}
  decr :: RIO <pTrue, decCtr> ()
  
  decCtr = \w _ w' -> 
    w' == update w ctr ((select ctr w) - 1)
\end{code}
\toolname verifies that at the end of @whileTest@ 
the counter is zero (\ie the post-condition @zeroCtr@)
by instantiating suitable (\ie inductive) refinements
for this particular use of @whileM@.
 



\section{Capability Safe Scripting via \RIO}
\label{sec:files}\label{subsec:state:files}
\begin{figure}[t]
\begin{mcode}
pread, pwrite, plookup, pcontents,
pcreateD, pcreateF, pcreateFP :: Priv -> Bool

active   :: World -> Set FH 
caps     :: World -> Map FH Priv

pset p h = \w -> p (select (caps w) h) && 
                 h $\in$ active w
\end{mcode}
\caption{\label{fig:fstypes} Privilege Specification}
\end{figure}



Next, we describe how we use the \RIO monad to reason about shell
scripting, inspired by the \shill~\citep{shill} programming language.

\paragraph{\shill} is a scripting language that restricts the
privileges with which a script may execute by using
\emph{capabilities} and \emph{dynamic contract checking}~\citep{shill} .
Capabilities are \emph{run-time values} 
that witness the right to use a particular resource 
(\eg a file).
A capability is associated with a set of privileges, 
each denoting the permission to use the capability 
in a particular way (such as the permission to write 
to a file).
A contract for a \shill procedure describes the 
required input capabilities and any output values.
The \shill runtime guarantees that system resources are accessed in
the manner described by its contract.

In this section, we turn to the problem of
preventing \shill runtime failures.
(In general, the verification of file system resource usage is a rich
topic outside the scope of this paper.)
That is, assuming the \shill runtime and an API as described in
\cite{shill}, how can we use Bounded Refinement Types to encode
scripting privileges and reason about them \emph{statically?}

\paragraph{We use \RIO types} to specify \shill 's API operations
thereby providing \emph{compile-time} guarantees 
about privilege and resource usage.
To achieve this, we:
connect the state (@World@) of the \RIO monad with a
\emph{privilege specification} denoting the set of 
privileges that a program may use~(\S~\ref{sec:privilege-spec});
specify the \emph{file system API} in terms of this
abstraction~(\S~\ref{sec:fs-api});
and use the above to specify and verify the particular 
privileges that a \emph{client} of the API uses~(\S~\ref{sec:fs-client}).

\subsection{Privilege Specification}
\label{sec:privilege-spec}
Figure~\ref{fig:fstypes} summarizes how we specify privileges 
inside @RIO@. 
We use the type @FH@ to denote a file handles, analogous to \shill's
capabilities. An abstract type @Priv@ denotes the sets of privileges
that may be associated with a particular @FH@.

\paragraph{To connect \texttt{World}s with Privileges} we assume 
a set of uninterpreted functions of type @Priv ->  Bool@ 
that act as predicates on values of type @Priv@, each 
denoting a particular privilege.
For example, given a value @p :: Priv@, the proposition 
@pread p@ denotes that @p@ includes the ``read'' privilege.
The function @caps@ associates each @World@ with a @Map FH Priv@,
a table that associates each @FH@ with its privileges.
The function @active@ maps each @World@ to the @Set@ of
allocated @FH@s.
Given @x:FH@ and @w:World@, @pwrite (select (caps w) x)@
denotes that in the state @w@, the file @x@ 
may be written.
This pattern is generalized by the predicate @pset pwrite x w@.

\subsection{File System API Specification}
\label{sec:fs-api}
A privilege tracking file system API can be partitioned into the
privilege \emph{preserving} operations and the privilege \emph{extending}
operations.

\paragraph{To type the privilege preserving} operations, we define a predicate
@eqP w w'@ that says that the set of privileges and active handles
in worlds @w@ and @w'@ are \emph{equivalent}.
\begin{code}
  eqP = \w _ w' -> 
    caps w == caps w' && active w == active w'
\end{code}
We can now specify the privilege preserving operations that @read@ and @write@ files, 
and list the @contents@ of a directory, all of which require the 
capabilities to do so in their pre-conditions:
\begin{code}
  read :: {- Read the contents of h -}
    h:FH -> RIO<pset pread h, eqp> String
  
  write :: {- Write to the file h -}
    h:FH -> String -> RIO<pset pwrite h, eqp> ()
  
  contents :: {- List the children of h -}
    h:FH -> RIO<pset pcontents h, eqp> [Path]
\end{code} 

\paragraph{To type the privilege extending} operations, we define 
predicates that say that the output world is suitably 
extended. First, each such operation \emph{allocates} 
a new handle, which is formalized as:
\begin{mcode}
  alloc w' w x = 
    (x $\not \in$ active w) && active w' == {x} $\cup$ active w
\end{mcode}
which says that the active handles in (the new @World@) 
@w'@ are those of (the old @World@) @w@ extended with the
hitherto \emph{inactive} handle @x@.
Typically, after allocating a new handle, a script will
want to add privileges to the handle that are obtained
from existing privileges.

\paragraph{To \texttt{create} a new file} in a directory with handle @h@ we
want the new file to have the privileges \emph{derived} from
@pcreateFP (select (caps w) h)@ (\ie the create privileges of @h@). We
formalize this by defining the post-condition of @create@ as the predicate @derivP@:
\begin{code}
  derivP h  = \w x w' -> 
    alloc w' w x && 
    caps w' == store (caps w) x 
                  (pcreateFP (select (caps w)) h)

  create :: {- Create a file -}
    h:FH->Path->RIO<pset pcreateF h, derivP h> FH
\end{code}
Thus, if @h@ is writable in the old @World w@ 
(@pwrite (pcreateFP (select (caps w) h))@) and
@x@ is derived from @h@ (@derivP w' w x h@ both hold),
then we know that @x@ is writable in the new @World w'@
(@pwrite (select (caps w') x)@).

\paragraph{To \texttt{lookup} existing files} or create sub-directories,
we want to directly \emph{copy} the privileges of the parent handle. 
We do this by using a predicate @copyP@ as the post-condition for 
the two functions:
\begin{code}
  copyP h = \w x w' ->
    alloc w' w x && 
    caps w' == store (caps w) x 
                     (select (caps w) y)

  lookup :: {- Open a child of h -}
    h:FH->Path->RIO<pset plookup h, copyP h> FH

  createDir :: {- Create a directory -}
    h:FH->Path->RIO<pset pcreateD h, copyP h> FH
\end{code}
  
\subsection{Client Script Verification}
\label{sec:fs-client}
We now turn to a client script,
the program @copyRec@ 
that copies the contents of the directory @f@ to the
directory @d@.
\begin{code}
  copyRec recur s d = 
    do cs <- contents s
       forM_ cs $ \ p -> do
         x <- flookup s p
         when (isFile x) $ do
           y <- create d p
           s <- fread x
           write y s
         when (recur && (isDir x)) $ do
           y <- createDir d p
           copyRec recur x y
\end{code}
@copyRec@ executes by first listing the contents of @f@, 
and then opening each child path @p@ in @f@. 
If the result is a file, it is copied to the directory @d@.
Otherwise, @copyRec@ recurses on @p@, if @recur@ is true.

In a first attempt to type @copyRec@ we give it the following type:
\begin{code}
  copyRec :: Bool -> s:FH -> d:FH ->
             RIO<copySpec s d,
                 \_ _ w -> copySpec s d w> () 

 copySpec h d = \w ->
   pset pcontents h w && pset plookup h     w &&
   pset pread h     w && pset pcreateFile d w &&
   pset pwrite d    w && pset pcreateF d    w &&
   pwrite (pcreateFP (select (caps w) d)))
\end{code}              
The above specification gives @copyRec@ a minimal set of privileges. 
Given a source directory handle @s@ and destination handle @d@, the
@copyRec@ must at least:
\begin{inparaenum}[(1)]
  \item list the contents of @s@ (@pcontents@),
  \item open children of @s@ (@plookup@),
  \item read from children of @s@ (@pread@),
  \item create directories in @d@ (@pcreateD@),
  \item create files in @d@ (@pcreateF@), an
  \item write to (created) files in @d@ (@pwrite@).
\end{inparaenum}
Furthermore, we want to restrict the privileges on newly created files
to the write privilege, since @copyRec@ does not need to read from or
otherwise modify these files.

Even though the above type is sufficient to verify
the various clients of @copySpec@ it
is insufficient to verify @copySpec@'s implementation, 
as the postcondition merely states that @copySpec s d w@ holds.
Looking at the recursive call in the last line of @copySpec@'s implementation,
the output world @w@ is only known to satisfy @copySpec x y w@ (having
substituted the formal parameters @s@ and @d@ with the actual @x@ and
@y@), with no mention of @s@ or @d@!
Thus, it is impossible to satisfy the postcondition of @copyRec@, as
information about @s@ and @d@ has been lost.

\paragraph{Framing} is introduced to address the above problem.
Intuitively, because no privileges are ever
\emph{revoked}, if a privilege for a file existed \emph{before} the
recursive call, then it exists \emph{after} as well.
We thus introduce a notion of \emph{framing} -- assertions about
unmodified state that hold before calling @copyRec@ must hold after
@copyRec@ returns.
Solidifying this intuition, we define a predicate @i@ to be @Stable@
when assuming that the predicate @i@
holds on @w@, if @i@ only depends on the allocated set of 
privileges, then @i@ will hold on a world @w'@ so long as
the set of priviliges in @w'@ contains those in @w@.
The definition of @Stable@ is derived precisely from the ways in which
the file system API may modify the current set of privileges:
\begin{code}
  bound Stable i = \x y w w' -> 
   i w => ( eqP w () w' || copyP y w x w'
           || derivP y w x w'
          ) => i w'
\end{code}
We thus parameterize @copyRec@ by a predicate @i@, bounded by @Stable i@, 
which precisely describes the possible world transformations under which 
@i@ should be stable:
\begin{code}
  copyFrame i s d = \w -> i w && copySpec s d w

  copyRec :: (Stable i) => 
             Bool -> s:FH -> d:FH ->
             RIO<copyFrame i s d,
                 \_ _ w -> copyFrame i s d w> () 
\end{code}              
Now, we can verify @copyRec@'s body, as
at the recursive call that appears in the last line of the implementation,
@i@ is instantiated with 
@\w -> copySpec s d w@.

\section{Related Work}\label{sec:related}


\paragraph{Higher order Logics and Dependent Type Systems}
including
NuPRL~\citep{Constable86},
Coq~\citep{coq-book}, Agda~\citep{norell07},
and even to some extent, \haskell~\citep{JonesVWW06, McBride02},
occupy the maximal extreme of the expressiveness spectrum.
However, in these settings, checking requires explicit
proof terms which can add considerable programmer overhead.
Our goal is to eliminate the programmer overhead of
proof construction by restricting specifications to
decidable, first order logics and to see how far
we can go without giving up on expressiveness.
The \fstar system enables full dependent typing via
SMT solvers via a higher-order universally quantified
logic that permit specifications similar to ours
(\eg @compose@, @filter@ and @foldr@).
%
While this approach is at least as expressive
as bounded refinements it has two drawbacks.
First, due to the quantifiers, the generated VCs
fall outside the SMT decidable theories.
This renders the type system undecidable (in theory),
forcing a dependency on the solver's unpredictable
quantifier instantiation heuristics (in practice).
Second, more importantly, 
the higher order
predicates must be \emph{explicitly} instantiated,
placing a heavy annotation burden on the programmer.
In contrast, bounds permit decidable
checking, and are automatically instantiated
via Liquid Types.

\paragraph{Our notion of Refinement Types}
has its roots in the predicate subtyping
of PVS~\cite{Rushby98} and \emph{indexed types}
(DML~\cite{pfenningxi98}) where types are constrained
by predicates drawn from a logic.
To ensure decidable checking several refinement
type systems including~\citep{pfenningxi98,Dunfield07,LiquidICFP14}
restrict refinements to decidable, quantifier free logics.
While this ensures predictable checking and inference~\cite{LiquidPLDI08}
it severely limits the language of specifications, and makes it hard to
fashion simple higher order abstractions like @filter@ (let alone the more
complex ones like relational algebras and state transformers.)


\paragraph{To Reconcile Expressiveness and Decidability}
\catalyst~\citep{catalyst} permits a form of
higher order specifications where refinements
are relations which may themselves be parameterized
by other relations, which allows for example, a
way to precisely type @filter@ by suitably
composing relations.
However, to ensure decidable checking, \catalyst
is limited to relations that can be specified as
catamorphisms over inductive types, precluding
for example, theories like arithmetic.
More importantly, (like \fstar), \catalyst provides
no inference: higher order relations must be
\emph{explicitly} instantiated.
Bounded refinements build directly upon
abstract refinements~\citep{vazou13},
a form of refinement polymorphism
analogous to parametric polymorphism.
While \cite{vazou13} adds expressiveness via
abstract refinements, without bounds we cannot
specify any \emph{relationships between} the
abstract refinements. The addition of bounds
makes it possible to specify and verify the examples
shown in this paper,
while preserving decidability and inference.

\paragraph{Our Relational Algebra Library} builds on a long
line of work on type safe database access.
The HaskellDB~\citep{haskellDB}
showed how phantom types could be used to eliminate
certain classes of errors.
Haskell's HList library~\citep{heterogeneous}
extends this work with type-level computation
features to encode heterogeneous lists, which
can be used to encode database schema, and
(unlike HaskellDB) statically reject accesses
of ``missing'' fields.
The HList implementation is non-trivial,
requiring new type-classes for new operations
(\eg @append@ing lists); \citep{thepipower}
shows how a dependently typed language greatly
simplifies the implementation.
Much of this simplicity can be recovered in
Haskell using the @singleton@ library~\citep{Weirich12}.
Our goal is to show that bounded refinements
are expressive enough to permit the construction
of rich abstractions like a relational algebra
and generic combinators for safe database access
while using SMT solvers to provide decidable
checking and inference. Further, unlike the
HList based approaches, refinements they can
be used to \emph{retroactively} or \emph{gradually}
verify safety; if we erase the types we still
get a valid Haskell program operating over
homogeneous lists.


\paragraph{Our Approach for Verifying Stateful Computations} using monads
indexed by pre- and post-conditions is inspired by the method of
Filli\^atre~\citep{Filliatre98}, which was later enriched with
separation logic in Ynot~\citep{ynot}. In future work it would
be interesting to use separation logic based refinements to specify
and verify the complex sharing and aliasing patterns allowed by Ynot.
\fstar encodes stateful computations in a special Dijkstra
Monad~\citep{dijkstramonad} that replaces the two assertions with
a single (weakest-precondition) predicate transformer which
can be composed across sub-computations to yield a transformer
for the entire computation.
Our \RIO approach uses the idea of indexed monads but
has two concrete advantages.
First, we show how bounded refinements alone suffice to
let us fashion the \RIO abstraction from scratch.
Consequently, second, we automate inference of pre- and
post-conditions and loop invariants as refinement instantiation
via Liquid Typing.

\subsection*{Acknowledgments}
We thank the anonymous reviewers and Colin Gordon for providing invaluable feedback
about earlier drafts of this paper.

{
\bibliographystyle{plain}
\bibliography{sw}
}

\end{document}